\documentclass[floatfix,reprint,aps,prc,showpacs,showkeys,eqsecnum,amsmath,amsfonts,amssymb,
twocolumn,groupedaddress,superscriptaddress,nofootinbib,preprintnumbers]{revtex4-1}
\usepackage{graphicx}
\usepackage{dcolumn}
\usepackage{bm}
\usepackage{hyperref} 
\usepackage{float} 

\usepackage{amsmath} 
\usepackage{amssymb} 
\usepackage{amsfonts} 
\usepackage{latexsym}
\usepackage{enumitem}

\usepackage[english]{babel}
\usepackage[expansion=all,babel]{microtype}

\usepackage[usenames]{xcolor}

\newcommand{\be}{\begin{equation}}
\newcommand{\ee}{\end{equation}}
\newcommand{\ba}{\begin{eqnarray}}
\newcommand{\ea}{\end{eqnarray}}
   
\newcommand{\Singlet}{\mbox{$^1$S$_0$} \,}
\newcommand{\triplet}{\mbox{$^3$P$_2$} \,}
\newcommand{\Triplet}{\mbox{$^3$P-F$_2$} \,}

\newcommand{\Rs}{\mbox{$R_\sigma \,$}}
\newcommand{\Tcp}{\mbox{$T_{c, p}^\mathrm{max} \,$}}
\newcommand{\Tcn}{\mbox{$T_{c, n}^\mathrm{max} \,$}}
\newcommand{\Te}{\mbox{$T_\mathrm{eff} \,$}}

\newcommand{\Msun}{\mbox{$\mathrm{M}_\odot \,$}}

\newcommand{\XMMU}{J1732}

\newcommand{\dd}{\mathrm{d}}

\makeatletter
	\newcommand{\vast}{\bBigg@{2.85}}
\makeatother

\newcommand{\eq}[1]{Eq.~\eqref{#1}}
\newcommand{\fig}[1]{Fig.~\ref{#1}}

\newcommand{\sect}[1]{Section~\ref{#1}}

\newcommand{\aap}{Astron. Astrophys.}
\newcommand{\ssr}{Space Science Reviews}

\begin{document}

\preprint{INT-PUB-18-029}

\title{Constraints on Axion-like Particles and Nucleon Pairing in Dense Matter \\
from the Hot Neutron Star in HESS J1731-347}
\author{Mikhail V. Beznogov}
\email{mikhail@astro.unam.mx}
\affiliation{Instituto de Astronom\'ia, Universidad Nacional Aut\'onoma de M\'exico, Ciudad de  M\'exico, 04510, Mexico}
\author{Ermal Rrapaj}
\email{ermalrrapaj@gmail.com}
\affiliation{Department of Physics, University of Guelph, Ontario N1G 2W1, Canada}
\author{Dany Page}
\email{page@astro.unam.mx}
\affiliation{Instituto de Astronom\'ia, Universidad Nacional Aut\'onoma de M\'exico, Ciudad de  M\'exico, 04510, Mexico}
\author{Sanjay Reddy}
\email{sareddy@uw.edu}
\affiliation{Institute for Nuclear Theory and Department of Physics, University of Washington, Seattle,  WA 98195. }

\date{\today}

\begin{abstract}
If the thermal evolution of the hot young neutron star in the supernova remnant HESS J1731-347 is driven by neutrino emission, 
it provides a stringent constraint on the coupling of light (mass~$\ll 10$~keV) axion-like particles to neutrons. 
Using Markov-Chain Monte Carlo we find that for the values of axion-neutron coupling $g_{ann}^2 > 7.7 \times 10^{-20}$ (90\% c.l.) the axion cooling from the bremsstrahlung 
reaction $n+n\rightarrow n+n +a$ is too rapid to account for the high observed surface temperature. 
This implies that the Pecci-Quinn scale or axion decay constant $f_a > 6.7 \times 10^7$ GeV for KSVZ axions and $f_a > 1.7 \times 10^9$ GeV for DFSZ axions. The high temperature of this neutron star also allows us to tighten constraints on the size of the nucleon pairing gaps.
\end{abstract}


\maketitle

\section{Introduction}
\label{sec:intro}

The neutron star XMMU J173203.3--344518, hereafter \XMMU, was discovered in X-ray observations by Suzaku, XMM-Newton, 
and Chandra \cite{Acero:2009fk,Tian:2010uq} as a point source in the center of the shell-type supernova remnant G353.6-0.7
which itself was found \cite{Tian:2008kx} coincident with the high energy source HESS J1731-347
discovered in a Galactic plane survey with the H.E.S.S. observatory \cite{Aharonian:2008vn}.
Neither a radio pulsar nor an extended radio or X-ray counterpart that could be identified as a pulsar wind nebula, 
and revealing the presence of a pulsar whose beam is not pointing toward us \cite{Gaensler:2006aa}, has been found coincident with this source \cite{Tian:2008kx}.
These characteristics allow to classify this neutron star (NS) as a Central Compact Object (CCO: \cite{Pavlov:2002fr}),
a new member of a growing family of isolated neutron stars found in the center of young supernova remnants that present no, or little, evidence
for the presence of a significant magnetic field (see, e.g., \cite{Gotthelf:2008mz,de-Luca:2008gf,Potekhin:2015ul,De-Luca:2017pd} for reviews).
J1732 is hence a young, age $\simeq 30$ kyr, neutron star with a high surface temperature, $\Te\simeq 2.2 \times 10^6$ K 
\cite{Klochkov:2015rr,Ofengeim:2015daa} that makes it the hottest known cooling neutron star, with likely a very weak magnetic field.
(Young magnetars can be hotter: they are, however, not considered as passive coolers but rather being maintained hot by the decay of their
intense internal magnetic fields, see, e.g., \cite{Potekhin:2015wd}.)

Young isolated cooling neutron stars have their temperature evolution driven by neutrino emission from birth, in a supernova explosion,
till they reach an age of the order of a hundred thousands of years when photon emission from their hot surface becomes dominant
(see, e.g., \cite{Yakovlev:2004iq,Page:2006aa,Potekhin:2015wd} for reviews).
Neutrino emission itself is strongly altered by the occurrence of pairing since the development of a gap in the single particle excitation spectrum
can strongly suppress these excitations and, consequently, all neutrino emission processes in which the paired component participates
(see, e.g., \cite{Yakovlev:2001uq,Page2014} for reviews).
By sampling published results on neutron star cooling it is clear that only models with the lowest possible neutrino emission can 
explain such a high surface temperature as found for J1732.
Such low neutrino emission implies a minimal cooling scenario \cite{Page:2004fy} in which all fast neutrino process are {\em a priori} excluded:
results in, e.g., Figure 28 of \cite{Page:2004fy} show that models with no neutron pairing, but strong proton superconductivity, may be compatible with J1732.
In such a case both the modified Urca processes, 
$n+n' \rightarrow p+n' +e^-+\bar{\nu}_e$ and its inverse $p+n'+e^- \rightarrow n+n' + {\nu_e}$ in the neutron branch as well as 
$n+p' \rightarrow p+p' +e^-+\bar{\nu}_e$ and its inverse $p+p'+e^- \rightarrow n+p' + {\nu_e}$ in the proton branch,
and the two bremsstrahlung processes
$n+p\rightarrow n+p +\nu+\bar{\nu}$ as well as $p+p\rightarrow p+p +\nu+\bar{\nu}$,
that provide the basic neutrino emission in the standard cooling are strongly suppressed since they all involve protons
and only the neutron-neutron bremsstrahlung, $n+n\rightarrow n+n +\nu+\bar{\nu}$, is freely operating.
Beside the suppression of excitations and neutrino emission, pairing also triggers a new neutrino emission mechanism through the constant
formation and breaking of Cooper pairs when the development of the condensate is progressing at temperatures only slightly lower than the pairing
critical temperature $T_c$ \cite{Flowers:1976aa,Voskresenskii:1986aa},
a process that has been dubbed as ``PBF'' for Pair Breaking and Formation.
(When $T \ll T_c$ the PBF process is also strongly suppressed as all other neutrino process to which the paired component participate.)
This process is very efficient in the case of neutron \Triplet pairing in spin triplet but is suppressed for isotropic pairing as the neutron and proton \Singlet in spin singlets \cite{Leinson:2006fk}:
this difference is the reason why the lowest possible neutrino emission is obtained in the presence of only proton \Singlet superconductivity 
and the avoidance of neutron \Triplet superfluidity.
Neutron \Singlet superfluidity is only present at lower densities in the inner crust of the neutron star and has very little impact on the
neutrino emission.
Beside the presence of strong proton superconductivity the models also require the presence of light elements in the outer layers of the stars.
These layers provide a thermal insulating blanket, called the {\em envelope}, between the hot interior, whose temperature is determined by 
the neutrino losses, and the much colder surface: a light element envelope has a larger thermal conductivity  and results in a warmer surface 
while a heavy element envelope would result in a surface temperature much lower than observed in J1732.
Such models were initially explored in \cite{Klochkov:2015rr} and \cite{Ofengeim:2015daa} 

 Any process that emits a $\nu \overline{\nu}$ pair can also emit an axion.
 However, the unique conditions encountered in  J1732 require a strong suppression of all neutrino processes except the nn-bremsstrahlung.
 Consequently, axion emission by any of these suppressed processes can be neglected and we focus exclusively on axion emission due to the nn-bremsstrahlung reaction,
$n+n\rightarrow n+n + a$, which would operate if axions exist and couple to neutrons.
A constraint on the axion emissivity (energy loss/volume/time) can be used to constrain the axion-neutron coupling constant 
$g_{ann} = c_n m_n/f_a$  where $f_a$ is the axion decay constant, $m_n$ is the mass of the neutron, and $c_n$ is a model-dependent dimensionless number.    
We show in \S~\ref{sec:axion_emission}
that both the standard neutrino emission rate, and the axion emission rate depend on the same underlying nuclear physics and are determined by the rate of neutron spin fluctuations in the dense medium. 
Thus, if the cooling of J1732 is dominated by neutron bremsstrahlung processes, the observed temperature constrains $g_{ann}$ rather uniquely.   
Since there is a large uncertainty on the estimate of this neutron star age, and no direct constraint on its mass
we employ fiducial values, and error estimates and by means of Markov-Chain Monte Carlo (MCMC) understand model systematics and error propagation to provide an improved constraint on $g_{ann}$. 
We shall compare this constraint to the widely discussed constraint on $g_{ann}$ and $g_{app}$ obtained earlier using the duration of the neutrino signal observed from supernova (SN) 1987A more than thirty years ago \cite{Koshiba1992} and find that our constraint for $g_{ann}$ is competitive with the supernova bounds.

In section \ref{sec:prem} we provide a brief description of axions and their coupling to nucleons.
In section \ref{sec:axion_emission} we compare the neutrino and axion emission rates and verify that the axion once emitted,
free streams out of the star. 
We describe the family of neutron star cooling scenarios that would be compatible with observed high $T_\mathrm{eff}$ of J1732  in
section \ref{sec:setup} and characterize it in details in the next section \ref{sec:basic}.
We finally apply this characterization to obtain our new constraint on $g_{ann}$ in the section \ref{sec:axion_constraint}
and discuss the relation of our constraints on $g_{ann}$ and on nucleon pairing with other works in \S~\ref{sec:discussion}.
We conclude in section \ref{sec:conclusion} where we also comment on the implications from future observations of young and hot neutron stars.

\section{Axions and earlier constraints}
\label{sec:prem}

Axions arise as Goldstone bosons associated with the breaking of a global chiral $U(1)$ symmetry introduced by Peccei and Quinn \cite{Weinberg1978,Wilczek1978}. This symmetry was introduced to provide a natural explanation to the smallness of the CP violating $\theta-$term in QCD and the axion remains the best motivated weakly interacting light particle beyond the standard model.  To resolve the strong CP-problem the standard model Lagrangian is extended by introducing a coupling between axions and gluons. This coupling is uniquely specified by one parameter, $f_a$, with the dimension of energy,  and is described by the Lagrangian density
\begin{equation}
 \begin{split}
  \mathcal{L}=&\frac{g_s}{32 \pi^2}\frac{a}{f_a}G^b_{\mu \nu}\tilde{G}^{b}{}^{\mu \nu}
 \end{split}
\end{equation}
where $a$ is the axion field, ($\tilde{G}$) $G$  is the (dual) gluon field with $b$ as the color index; $g_s$ is the QCD coupling constant.
At low energies the Pecci-Quinn symmetry (PQ) is spontaneously broken, and the axion acquires a mass
\begin{equation}
  m_a=\frac{z^{1/2}}{1+z}\frac{f_{\pi}m_{\pi}}{f_a} \; ,
 \label{eq:QCDaxion}
\end{equation}
where $z=m_u/m_d\approx 0.56$ \cite{Gasser198277,Leutwyler:1996qg}. 
The axion coupling to the nucleon axial current at low energy is described by the Lagrangian
\begin{equation}
{\cal L}_{aN}=\frac{c_N}{2f_a}~\partial_\mu a~\bar{N}\gamma^\mu \gamma_5N, 
\label{eq:axion-neutron}
\end{equation} 
where $N=n,p$ is the nucleon field and $c_N$ is a model dependent dimensionless constant which, as noted before, is related to the coupling strength $g_{aNN}= c_N m_n/f_a$. 
In the KSVZ-like models the axions do not couple to the quark axial currents and a recent determination using lattice QCD finds that the gluonic contribution gives $c_p^{\rm KSVZ}=-0.47(3)$ and $c_n^{\rm KSVZ}=-0.02(3)$ \cite{diCortona:2015ldu}.
In the DFSZ model, where the Higgs sector contains two scalar fields, the axion couples to the quark axial currents. 
In this case, lattice calculations, also reported in Ref.~\cite{diCortona:2015ldu}, find 
\begin{align} 
c_n^{\rm DFSZ} &= 0.254 -0.414 \sin^2{\beta}\pm 0.025\, \\
c_p^{\rm DFSZ} &= -0.617 + 0.435 \sin^2{\beta}\pm 0.025\,,  
\end{align} 
where $\cot{\beta}$ is the ratio of the vacuum expectation values of the two Higgs fields in the DFSZ model. Axion-like particles (ALPs) also appear in the string theories into which the standard model can be embedded. In these models $m_a$ and $f_a$ are unrelated and Eq.~\ref{eq:QCDaxion} does not hold, but otherwise, these ALPs couple to electromagnetic and matter fields in a manner similar to the QCD axion. Since we shall focus on axion production from neutron-neutron bremsstrahlung reactions, our analysis and the constraints we derive for $g_{ann}$ will apply also to ALPs which couple to neutrons through the low-energy Lagrangian in Eq.~\ref{eq:axion-neutron}.

Astrophysical constraints on the axion-nucleon coupling have been discussed extensively in the literature, and Refs.~\cite{Raffelt:2006cw,Giannotti:2017hny} provide succinct reviews of the current state of the art. The most stringent constraint on the axion coupling to neutron and protons is derived from the duration of the neutrino signal observed from SN 1987A \cite{Raffelt:2006cw}. In the standard scenario, core-collapse supernovae produce a proto-neutron star (PNS) which cools and loses lepton number over a time scale of several tens of seconds as neutrinos diffuse through  hot dense matter encountered in the PNS \cite{Burrows:1986me}. During this time the PNS emits neutrinos copiously and the neutrino luminosity and typical energies are large enough to be detected in terrestrial neutrino detectors. Approximately 20 neutrino events detected from the supernova in 1987 (SN 1987A) over several seconds in Kamiokande and Irvine-Michigan-Brookhaven (IMB) neutrino detectors confirmed this basic picture \cite{Burrows:1987zz}. It was realized shortly afterwards that additional cooling due to axions would reduce the duration of the neutrino signal if the axion coupling was large enough to produce them effectively but not too large to ensure that the axion mean free path was larger than the PNS radius \cite{Burrows:1988ah}. Since the PNS contains partially degenerate matter with about $30-40$ \% protons during the early stages bremsstrahlung reactions $nn \rightarrow nn a$ and $pp \rightarrow pp a$ contribute to axion emission with approximately equal strengths. Although there remain important uncertainties associated with the calculations of the rate of axion bremsstrahlung of nucleons in a dense plasma, the ambient conditions encountered in the PNS and its early evolution, simulations and their analysis suggests 
\begin{equation} 
g^2_{ann} + g^2_{app} < 3.6 \times 10^{-19} 
\label{eq:sn1987a}
\end{equation} 
as an indicative rather than conclusive bound on the axion-nucleon couplings \cite{Patrignani:2016xqp}. For convenience we define $g^2_{aNN-87A}= 3.6 \times 10^{-19}$ and show that the bounds we obtain from J1732 are competitive. 

\section{Axion and neutrino emission}
\label{sec:axion_emission}

Both neutrino and axion emission arise due to nucleon spin fluctuations induced by nucleon-nucleon collisions \cite{Friman:1978zq}. 
The corresponding neutrino and axion emissivities due to the bremsstrahlung reaction have been calculated and can be written as  
\begin{equation}
\epsilon^{nn}_\nu=\frac{G^2_F}{60~\pi^4}~\int_0^\infty d\omega ~\omega^6~S_\sigma(\omega) \,,
\end{equation}
and 
\begin{equation}
\epsilon^{nn}_a=\frac{g^2_{ann}}{48~\pi^2m^2_n}~\int_0^\infty d\omega ~\omega^4~S_\sigma(\omega) \,,
\end{equation}
respectively. Here $G_F=1.166\times 10^{-5}$ GeV$^{-2}$ is the Fermi weak interaction constant. 
The function
%
\begin{align}
S_\sigma(\omega)=  \int \prod_{i=1}^{4}~\frac{d^3p_i}{(2\pi)^3}  (2\pi)^4\delta^3(\vec{p}_1+\vec{p}_2-\vec{p}_3+\vec{p}_4) \nonumber \\
\times \delta(E_1+E_2-E_3-E_4-\omega) ~f_1f_2\tilde{f}_3\tilde{f}_4~{\cal H}\,,
\label{eq:Ssigma}
\end{align}
is called the dynamic structure function and it characterizes the spectrum of spin fluctuations due to neutron-neutron collisions. Here ${\cal H}$ is the squared matrix element defined in Eq. 9 of Ref.~\cite{Hanhart:2000ae} and the other factors appearing in Eq.~\ref{eq:Ssigma} are the fermi functions $f_i=(1+\exp{(\beta(E_i-\mu_n)})^{-1}$ and $\tilde{f}_i=(1+\exp{(\beta(\mu_n-E_i)})^{-1}$, where $\mu_n$ is the neutron chemical potential and $\beta =1/kT$. The incoming nucleon momenta $p_1$ and $p_2$ and energies $E_1$ and $E_2$, outgoing nucleon momenta $p_3$ and $p_4$ and energies $E_3$ and $E_4$, and the total energy of the neutrino pair $\omega$. The matrix element squared ${\cal H}$ is a function of the nucleon momenta and the nucleon-nucleon interaction potential. Note that we do not include neutrino momenta in the momentum conserving delta function because the typical neutrino momenta are of order $p_\nu\approx T\simeq 10^{-2}$ MeV and are negligible compared to the typical momenta of neutrons at the fermi surface $p_i \simeq p_{Fn}=(3\pi^2 n_n)^{1/3} \simeq 331.4~(n_n/n_0)^{1/3}$ MeV where $n_n$ is the neutron number density and $n_0=0.16$ fm$^{-3}$ is the nuclear saturation density. 

Under degenerate conditions $\mu_N \gg T$  only neutrons at Fermi surface contribute, and further since $\omega\simeq T$ is small compared to all other energy scales associated with nucleons and the nucleon-nucleon interaction, the phase space integration can be done (Friman \& Maxwell 1979) and the dynamic structure factor has the general form given by 
\begin{equation}
S_{\rm \sigma}(\omega)=C_{\sigma} ~\frac{\omega^2+(2\pi T)^2}{\omega (\exp{(\beta \omega)-1)}}\,,
\label{eq:Ssigma2}
\end{equation}
where $C_{\sigma}$ depends on nucleon-nucleon interaction and is a weak function of the neutron density.  

In a simple model where neutrons interact due to the exchange of a single pion (in the Born approximation)\cite{Friman:1978zq}, 
\begin{equation}
{\cal H} = 8\left(\frac{f}{m_\pi}\right)^4 \left(\frac{k^2}{k^2+m_\pi^2}\right)^2~\frac{1}{\omega^2}\,,
\end{equation}
where $\vec{k}=\vec{p}_1-\vec{p}_3$ is the momentum transfer between neutrons.  In this model
\begin{equation}
C_\sigma \longrightarrow C_{\rm OPE} =\frac{2f^4M^4_N F(y)~p_{Fn}}{3\pi^5m_\pi^4}\,,
\label{eq:SOPE}
\end{equation}
with $y=m_\pi/2p_{Fn}$, and the function 
\begin{eqnarray}
F(y)= 3 - 5 y \arctan\left(\frac{1}{y}\right) + \frac{y^2}{1+y^2} + 
\nonumber
\\
\frac{y^2}{\sqrt{1+2y^2}} \arctan\left( \frac{\sqrt{1+2y^2}}{y^2} \right) \; .
\end{eqnarray}

Using the general form of the structure function given in Eq.~\ref{eq:Ssigma2} the neutrino and axion emissivities can be calculated. We find that
(in ergs cm$^{-3}$ s$^{-1}$)      
\begin{eqnarray}
\epsilon^{nn}_\nu = 1.7\times 10^{20}  F(y) R_\sigma \!
\left(\frac{m_n^*}{m_n}\right)^4 \!\! \left(\frac{n_n}{n_0}\right)^{1/3}T^8_9
\label{Eq:eps_nu}
\end{eqnarray}
and
\begin{eqnarray}
\epsilon^{nn}_a = 1.1\times 10^{20} F(y) R_\sigma R_A \!
\left(\frac{m_n^*}{m_n}\right)^4 \!\! \left(\frac{n_n}{n_0}\right)^{1/3}  \! T^6_9  \;\;\;\;
\label{Eq:eps_ax}
\end{eqnarray}
where 
\be
R_\sigma \equiv \frac{C_\sigma}{C_{\rm OPE}}   \;\;\;\; \mathrm{and}  \;\;\;\; R_A \equiv \frac{g^2_{ann}}{g^2_{aNN-87A}} \; .
\label{Eq:Rs_RA}
\ee
$R_\sigma$ incorporates the corrections to the rate that would arise with improved treatment of the nucleon-nucleon interaction that go beyond tree-level one pion exchange, and other corrections that arise due to many-body effects such as screening, and suppression of low energy bremsstrahlung from multiple scattering (Landau-Pomeranchuk-Migdal, or LPM, effect) \cite{Hanhart:2000ae}. 
 Calculations based on realistic nucleon-nucleon interactions which also go beyond the Born approximation by including a non-perturbative treatment of the two-nucleon scattering have been performed \cite{Hanhart:2000ae,Schwenk:2003pj} 
and find that $C_{\sigma}\simeq C_{\rm OPE}/4$ for the range of conditions encountered in neutron stars. 
 Thus, we will consider the value $\Rs = 0.25$ as our fiducial value.
The ratio $R_A$ controls the relative importance of axion cooling measured in terms of the suggested upper limit $g^2_{aNN-87A}= 3.6 \times 10^{-19}$ discussed earlier.

In our study we shall assume that axions produced in the core can freely escape without being reabsorbed in the neutron star. To ensure that this holds we calculated the mean free path of thermal axions with energy $E_a \simeq T$  due to the processes $a nn \rightarrow nn$ (inverse bremsstrahlung) and find that 
\begin{equation}
 \begin{split}
  \langle \lambda_a \rangle \approx& \frac{43.47}{\pi^4} \frac{m_n^2}{g^2_{aNN} ~ C_{\sigma} T^2}\,. 
 \end{split}
\label{eq:axion_mfp}
\end{equation}
An approximate condition for axion to free streaming can be defined using the optical depth
\begin{equation}
 \begin{split}
  \tau_a=&\int_{0}^{R_{\text{NS}}} \frac{dr}{\langle \lambda_a \rangle} < 3\,. 
 \end{split}
\label{eq:opticaldepth}
\end{equation}
Using Eq.~\ref{eq:axion_mfp} we can rewrite Eq.~\ref{eq:opticaldepth} as 
 \begin{equation}
 \begin{split}
  f_a>&\frac{|C_N| \pi^2}{11.42}T\sqrt{\ C_{\sigma} R_{\text{NS}}}\\
  \approx& 10^4 ~\text{GeV} ~\bigg(\frac{T}{10^8 ~\text{K}}\bigg) ~\sqrt{\frac{R_{\text{NS}}}{10 ~\text{km}}} \sqrt{\frac{4 C_{\sigma}}{C_{\rm OPE}}}\,, 
 \end{split}
\end{equation}
and suggests that for typical values of interest here ($f_a\simeq 10^7-10^9$ GeV) axion trapping is not an issue. 

\section{Setting up the scenario}
\label{sec:setup}

An extensive comparison of the inferred high effective temperature of J1732 with neutron star cooling theory has been presented in
\cite{Klochkov:2015rr} and \cite{Ofengeim:2015daa}.
We here describe our scenario and explore the effects of its physical and astrophysical parameters to clearly characterize our basic setup.

Since the observed high \Te of J1732 excludes fast neutrino emission it excludes the presence of any form of ``exotic'' baryonic matter beyond just
neutrons and protons as any of these would invariably result in strong neutrino emission \cite{Yakovlev:2001uq,Page:2006ab}
and surface temperatures much below the observed one.
Our general scenario is, hence, a special case of the minimal cooling paradigm as described in \cite{Page:2004fy} and \cite{Page:2009aa}.
There is, thus, not much freedom in the choice of the core equation of state (EOS) and we build our neutron stars employing the dense matter EOS of \citet{Akmal:1998aa} (APR EOS thereafter).
With these restrictions the main effect of the choice of core EOS of our neutron stars is to determine 
the radius of the star for a given mass and the neutron/proton fractions and effective masses,
these being the main EOS dependent physical ingredients that play a role in the  cooling simulations.
Our treatment of neutrino emission and specific heat is as described previously in \cite{Page:2004fy} and \cite{Page:2009aa} 
with the change that for the emissivity of the $nn$-bremsstrahlung process we replace equation 28 in \cite{Page:2004fy} by \eq{Eq:eps_nu}.
The effect of adopting a different EOS would be almost exclusively covered in the different prediction for the neutron and proton
effective masses that appear linearly in the specific heat and quartically in the emissivities of \eq{Eq:eps_nu} and \eq{Eq:eps_ax}.
We will consider this possible model dependent small variability in our MCMC simulations
as well as quantify the effect of uncertainty on the neutron star mass in \sect{sec:basic}.

\subsection{Envelope and its Chemical Composition}
\label{sec:envelope}

\begin{figure}
	\includegraphics[width=1.0\columnwidth]{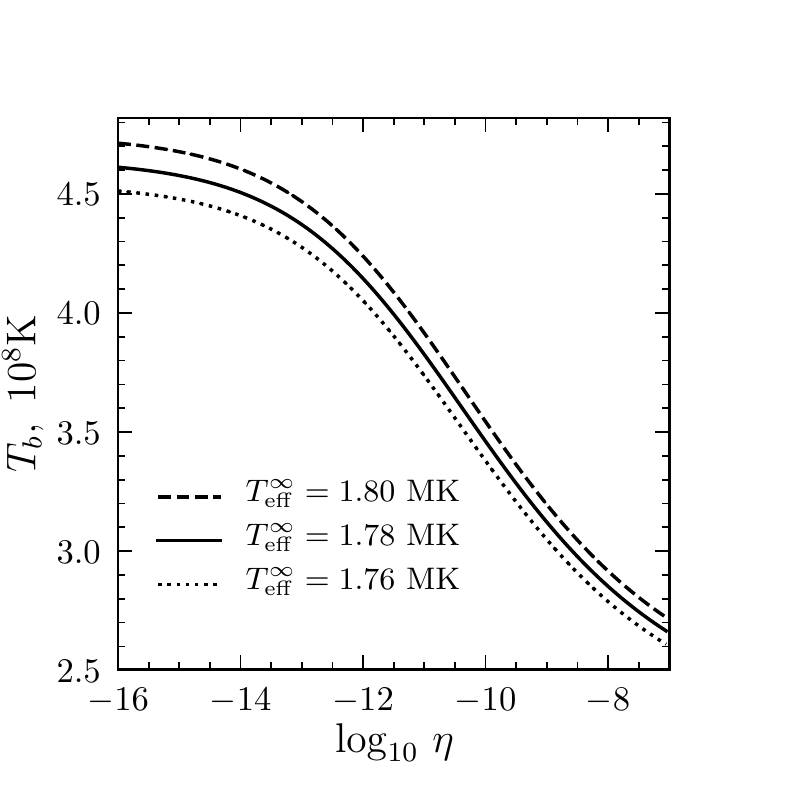}
	\caption{Interior temperature $T_b$ of J1732, as a function of the amount of carbon in the envelope, 
	parametrized by $\eta$ as in \eq{eq:eta}, for three effective temperatures.
	(From the envelope models of \cite{BPY16} and assuming a star of mass 1.4 \Msun with a radius of 11.8 km.)
	}
	\label{fig:eta}
\end{figure}

As shown in \cite{Klochkov:2015rr, Ofengeim:2015daa} J1732's soft X-ray spectrum is best characterized by a carbon atmosphere model. 
Thus, in all our calculations we assume that it has a carbon-iron heat blanketing envelope with a variable amount of carbon. In particular, 
we employ the fit for $T_\mathrm{eff} - T_b$ relation from \cite{BPY16}, $T_b$  being  the temperature 
at the bottom of the envelope at density $\rho_b = 10^{10}$~g~cm$^{-3}$. 
The amount of carbon in the envelope, $\Delta M_\mathrm{C}$, is parametrized by
\begin{equation}
\eta = g_{s14}^2 \frac{\Delta M_\mathrm{C}}{M}
\label{eq:eta}
\end{equation}
where $g_{s14}$ is the surface gravitational acceleration, in units of $10^{14}$ cm s$^{-2}$,  and $M$ is the neutron star mass.
The inferred interior temperature $T_b$ for an assumed amount of carbon is shown in \fig{fig:eta} for three effective temperatures that correspond
to the best fit and best fit $\pm$ error values of \Te for J1732 obtained by \cite{Klochkov:2015rr}.
The inference of presence of carbon at the surface from the spectral fit only means that the carbon layer is thick enough to ``hide from sight'' 
anything below it, i.e., its optical depth is larger than unity in the soft X-ray range, and only a few g cm$^{-2}$ of carbon are enough for this to occur.
We, thus, have no information about the actual thickness of this carbon layer and will consider it, i.e., $\eta$, as a free parameter
in our MCMC simulations.
We cannot exclude the presence of, and there certainly is, an intermediate layer composed of intermediate mass elements, O, Ne, Si, $\dots$,
between the carbon and the subjacent iron: its impact on the envelope insulating power is smaller than that of the carbon layer and is mocked up by 
an increase in $\eta$ and, hence, does not need to be explicitely taken into account.
The maximum amount of carbon we allow is $\eta \sim 10^{-7}$ and corresponds to carbon existing up to densities of the order of $3 \times 10^{9}$~g~cm$^{-3}$:
if pushed to higher densities carbon would have burned during the previous history of the star \cite{Yakovlev:1994xr}.
The smaller value we consider is $\eta \sim 10^{-16}$ since, as seen in \fig{fig:eta}, smaller value result in no change in $T_b$ compared to the $\eta = 10^{-16}$ case.

Notice that the star is old enough to have fully thermally relaxed and its red-shifted internal temperature $\widetilde T = e^\phi T$, at densities above $\rho_b$, 
is uniform and equal to $\widetilde T_b = e^{\phi_b} T_b$, where $e^\phi$ is the (density dependent) red-shift factor, i.e., the square root of the time-time component of the metric, $g_{00}$.

\subsection{Pairing}
\label{sec:pairing}

Given the constraints on possible models described above, a factor of major importance is the presence or absence of
pairing of either the neutrons or the protons in the neutron star core.
Nucleon pairing in neutron stars is predicted to occur either in the isotropic spin singlet \Singlet or the anisotropic spin triplet \Triplet phases,
the former at Fermi momenta $k_F$ below about $1 - 1.5$ fm$^{-1}$ and the latter at larger $k_F$.
For neutrons \Singlet pairing this corresponds to densities found in the neutron star inner crust of dripped neutrons and, at best, 
the outermost layers of the core,
while for protons it corresponds to densities found in the outer part of the core and possibly the whole core depending on the adopted 
pairing model and the neutron star mass. 
\Triplet pairing is expected for neutrons in the core and possibly protons at very high densities.
Theoretical calculations of the \Singlet gap sizes find maximum values around 1MeV, being about a factor of 3 smaller for protons than for neutrons,
but with large variations about the range of $k_F$ where the gap is significant.
In the case of the \Triplet gap for neutrons, some early calculations predicted value as large 1 MeV but presently predicted values are smaller, 
e.g., 0.1 MeV in \cite{Ding:2016aa} or even below 0.01 MeV in \cite{Schwenk:2004aa}.
We refer the reader to \cite{Page2014}, \cite{Gezerlis2014}, and \cite{Sedrakian:2018aa} for recent reviews. 

The \Singlet gap is nodeless, since it is spherically symmetric, while the \triplet gap can present many phases \cite{Zverev:2003aa}
and we consider only the so called ``B'' phase \cite{Levenfish:1994zp} with an angular dependence $1+3\cos \theta_k$ 
($\theta_k$ being the angle between the paired particle momentum $\bf k$ and some arbitrary quantization axis): 
this phase is also nodeless and found to be energetically favored.
Given the large theoretical uncertainties on both the size and the density dependence of the proton \Singlet and neutron \Triplet gap
we will explore their effects in our MCMC calculations.
In our cooling calculations the effect of pairing is entirely determined by the type of pairing, neutron or proton, and \Singlet or \triplet gaps,
and the value of the corresponding critical temperature $T_c$ which is density dependent.
For this purpose we model each gap's density dependence 
 through a dependence on the Fermi momentum $k_F$ of the neutron or proton with a Gaussian-like shape as
\ba
T_{c, N}(k_{F, N}) = 
\nonumber \\
T_{c, N}^\mathrm{max} \cdot \exp[-(k_{F, N} - k_{F, N}^\mathrm{max})^2/(\Delta k_{F, N})^2] \; , 
\label{eq:Tc}
\ea
for $N= n$ or $p$.
For our purpose the precise location of the peak is not very important as  long as it is well inside the core and we thus fix them at
$k_{F, p}^\mathrm{max} = 0.5 $ fm$^{-1}$ and $k_{F, n}^\mathrm{max} = 2.0 $ fm$^{-1}$.

\begin{figure}
	\includegraphics[width=1.0\columnwidth]{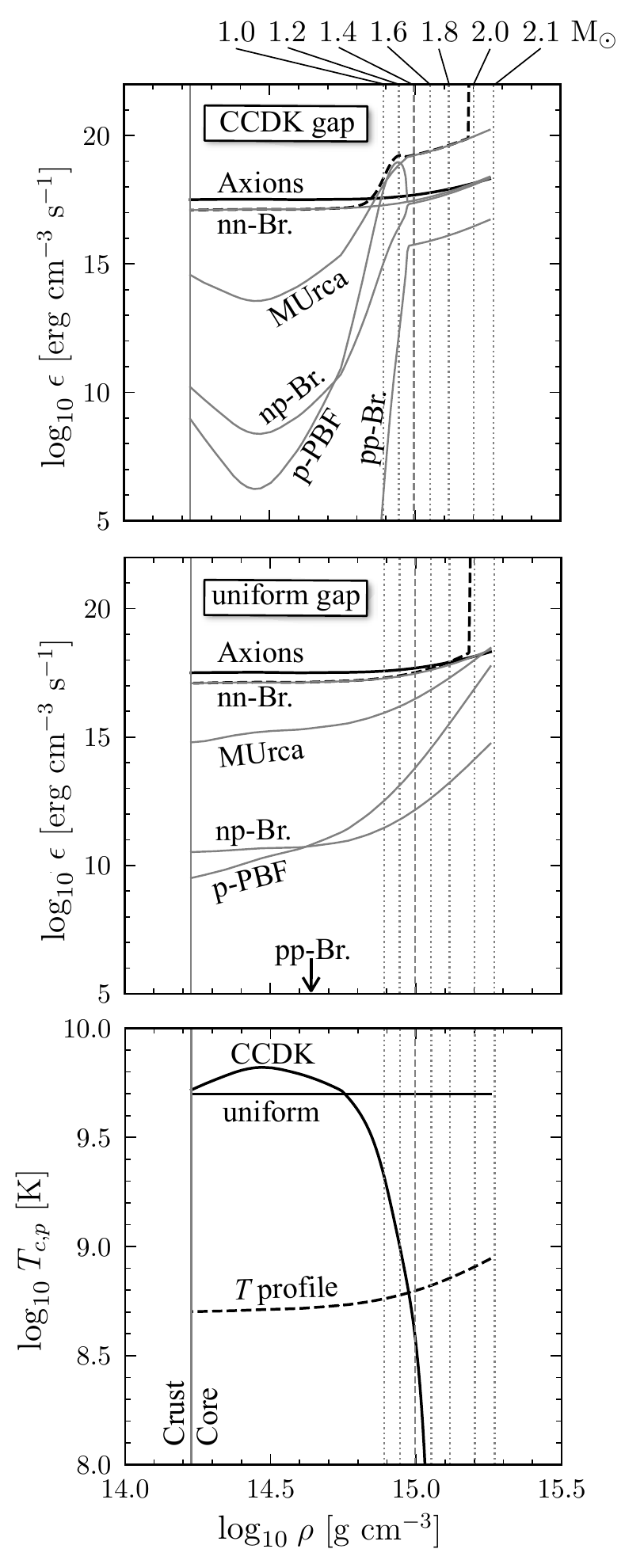}
	\caption{Examples of proton \Singlet pairing gaps, lower panel, and the resulting neutrino and axion emissivities,
	middle panel for uniform gap and upper panel for the CCDK \cite{Chen:1993fk} gap.
         See text for details.
	}
	\label{fig:Gap_nu_axion}
\end{figure}

As discussed in the introduction and in agreement with our findings in the next section,
neutron \Triplet pairing has to be avoided while proton \Singlet pairing has to be strong.
To illustrate the effect of proton superconductivity, in \fig{fig:Gap_nu_axion} we consider two distinct density dependencies for the proton \Singlet gap: a simple, albeit unrealistic, density independent gap, and density dependent one, ``CCDK'', taken from Ref.~\cite{Chen:1993fk} which makes a theoretical prediction that extends to the highest densities encountered.
The lower panel shows these two gaps and a temperature profile that is used to calculate neutrino and axion emissivities
which are plotted in the middle and upper panel where the uniform or CCDK gap, respectively, is assumed.
The emissivities correspond to a 2.1 \Msun star obtained with the APR EOS and the vertical dotted/dashed lines mark the central densities of lower mass stars with the same APR EOS, as labelled.
The solid vertical line indicates the crust-core boundary and the assumed temperature profile corresponds to an isothermal
interior with a uniform red-shifted temperature $\widetilde{T} = 10^{8.5}$~K
(which is representative of the recent interior temperature of J1732).
The plotted neutrino emissivities are: 
the modified Urca process, the sum of the neutron and proton branches, labelled as ``MUrca", 
the three bremsstrahlung processes labelled as ``nn-Br.", ``np-Br.", and ``pp-Br." and 
the PBF process from proton pairing labelled as ``p-PBF".
The total neutrino emissivity is shown by the dashed line which also exhibits the appearance of the direct Urca at densities
above $\approx 10^{15.2}$ g cm$^{-3}$.
The axion emissivity from nn-bremsstrahlung, \eq{Eq:eps_ax}, with $R_A=1$ is labelled as ``Axions".

In \fig{fig:Gap_nu_axion} one clearly sees the suppression of all process involving protons when $T \ll T_{c, p}$.
The nn-bremsstrahlung reaction dominates the total emissivity and axion emission competes with neutrinos for this specific choice of temperature and coupling constant.
In the case of the density dependent gap, upper panel, when $T$ approaches $T_{c, p}$ from below, at densities just below $10^{15}$ g cm$^{-3}$, both the modified Urca and
proton-PBF process rapidly increase and eventually dominate, resulting in a strong increase of the total neutrino emissivity.

\subsection{MCMC setup}
\label{sec:MCMC}

For our cooling calculations we use an up-dated version of the code {\tt NSCool} \cite{Page:2016zi} which solves the general relativistic equations
of energy balance and heat transport in spherical symmetry.
For the MCMC runs we use our driver {\tt MXMX} \cite{Page:2018ab} which is inspired by {\tt emcee} \cite{Foreman-Mackey:2013mg}.

We fit a cooling curve, i.e. a $T_\mathrm{eff}^\infty(t)$ function, to a best fit data point $(T_\mathrm{eff,obs}^\infty, t_\mathrm{obs})$
wich has uncertainties $\Delta T_\mathrm{eff,obs}^\infty$ and $\Delta t_\mathrm{obs}$.
We assume these observational values to be distributed according to a Gaussian and the uncertainty to be its standard deviation.
The likelihood in our case can be written, similarly to \cite{BHSP18}, in the following form (up to unimportant constants like normalization):
\begin{align}
	\begin{split}
		&L \sim \int \left[ \left(\frac{\dd T_\mathrm{eff}^\infty}{\Delta T_\mathrm{eff,obs}^\infty} \right)^2 + 
		                           \left(\frac{\dd t}{\Delta t_\mathrm{obs}} \right)^2      \right]^{1/2} \times\\
		&\exp{\vast[-\frac{\big(T_\mathrm{eff}^\infty - T_\mathrm{eff,obs}^\infty\big)^2}{2 \big(\Delta T_\mathrm{eff,obs}^\infty\big)^2} \vast]} 
		\exp{\vast[-\frac{\big(t - t_\mathrm{obs}\big)^2}{2 \big(\Delta t_\mathrm{obs}\big)^2} \vast]}
	\end{split}
\label{eq:Likelihood_int}	
\end{align}  
as a line integral along the cooling trajectory in the $t-T_\mathrm{eff}^\infty$ plane and where 
the units to measure $t$ and $T_\mathrm{eff}^\infty$ are naturally taken as the standard deviations for a Gaussian distribution in each axes, i.e., 
$\Delta t_\mathrm{obs}$ and $\Delta T_\mathrm{eff,obs}^\infty$.
As the cooling curves are computed at discretized time steps $t_i$, $i = 1, \ldots, i_\mathrm{max}$, we can replace integration by summation: 
\begin{align}
	\begin{split}
		&L  \sim \sum_{i=2}^{i_\mathrm{max}-1} \left[ \left(\frac{\Delta T_i}{\Delta T_\mathrm{eff,obs}^\infty} \right)^2 + 
		                                                                        \left(\frac{\Delta t_i}{\Delta t_\mathrm{obs}} \right)^2 \right]^{1/2}  \times\\
		&\exp{\vast[-\frac{\big(T_i - T_\mathrm{eff,obs}^\infty\big)^2}{2 \big(\Delta T_\mathrm{eff,obs}^\infty\big)^2} \vast]} 
		\exp{\vast[-\frac{\big(t_i - t_\mathrm{obs}\big)^2}{2 \big(\Delta t_\mathrm{obs}\big)^2} \vast]}, 
	\end{split}
\label{eq:Likelihood_sum}	
\end{align}  
where $T_i = T_\mathrm{eff}^\infty(t_i)$, $\Delta T_i = 0.5(T_{i+1} - T_{i-1})$ and $\Delta t_i = 0.5 (t_{i+1} -t_{i-1})$.

The study of J1732's soft X-ray thermal spectrum  in \cite{Klochkov:2015rr} showed that blackbody models are statistically rejected, i.e., give poor fits, 
while atmosphere models made of hydrogen or carbon both give good fits.
Hydrogen atmosphere models however imply too large a distance, above 7 kpcs, and high red-shifted effective temperatures, well above $2 \times 10^6$ K.
On the contrary, carbon atmosphere models imply a distance compatible with the estimated supernova remnant distance of 3--5 kpcs and a lower
red-shifted temperature around $2 \times 10^6$ K.
We accept, following \cite{Klochkov:2015rr}, that these carbon atmosphere models are the most credible ones and,
to be conservative we adopt the lowest values found by \cite{Klochkov:2015rr}, $T_\mathrm{eff,obs}^\infty = 1.78^{+0.04}_{-0.02}$ MK (redshifted effective surface 
temperature, 1$\sigma$ confidence level), which corresponds to the closest distance of 3.2 kpcs.
We will see that this value of 1.78 MK is already difficult to obtain in cooling models and very restrictive; 
adopting a larger value would only make all our conclusions stronger.
The age is highly uncertain; the typical age range mentioned in the literature is $10-40$ kyr and we adopt $t_\mathrm{obs} \sim 27$~kyr \cite{Tian:2008kx}
with 1$\sigma$ uncertainty $\Delta t_\mathrm{obs} = 6.5$~kyr.
For the effective temperature, since we want to be conservative in constraining axions, we use the smallest uncertainty of $\Delta T_\mathrm{eff,obs}^\infty = 0.02$ MK.

\section{Characterizing the Basic Scenario: Constraints on Nucleon Pairing}
\label{sec:basic}

\begin{figure}
	\includegraphics[width=1.0\columnwidth]{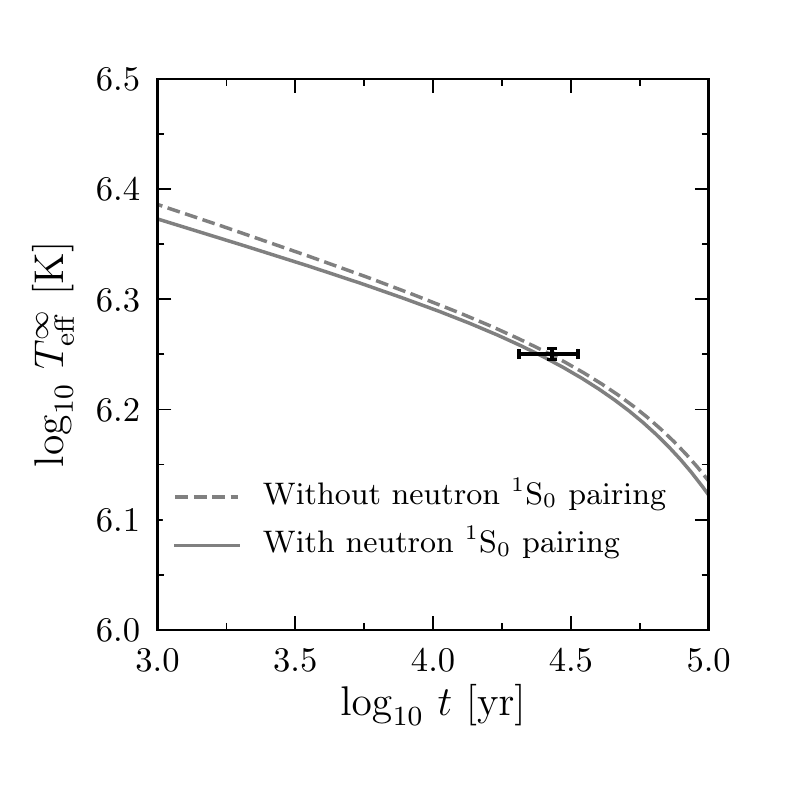}
	\caption{Cooling curves of effective temperature at infinity $T_\mathrm{eff}^\infty$ versus model age $t$
	         illustrating the effect of neutron \Singlet pairing (which is mostly located in the inner crust). 
		The data point corresponds to J1732 with $1\sigma$ uncertainties. 
		See text for details.}
	\label{fig:Cool_crustSF}
\end{figure}

Before embarking on the study of cooling scenario in the presence of axion emission we first characterize the basic scenario without axion.
We first consider the effect of neutron \Singlet superfluidity in the crust, then the effects of changing the neutron star model mass and finally,
in order to confirm, or infirm the claim of \cite{Klochkov:2015rr} and \cite{Ofengeim:2015daa} that neutron \Triplet superfluidity must be excluded and
that proton \Singlet superconductivity must be strong, we perform an MCMC study of it.

Starting with \Singlet neutron superfluidity we present in \fig{fig:Cool_crustSF} two illustrative cooling curves, one including it and the other excluding it,
all other parameters being identical.
We choose the model of \cite{Schwenk:2003vq} for $T_{c, n}$.
The difference between curves is simply due to the specific heat: pairing reduces specific heat of the crust and, thus, the star is cooling faster.
The difference between the two models is, however, quite small and employing different models for $T_{c, n}$ would produces cooling curves
with even smaller difference between them.
We will always include inner crust neutron superfluidity, taken from \cite{Schwenk:2003vq}, in our following calculations.

\begin{figure}
	\includegraphics[width=1.0\columnwidth]{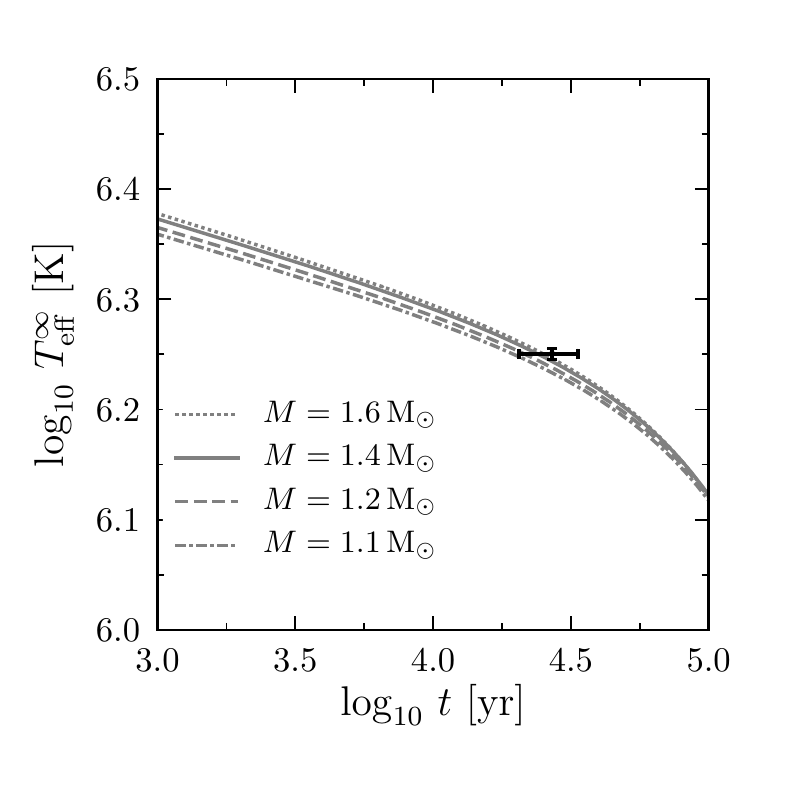}
	\caption{Same as \fig{fig:Cool_crustSF} with four models illustrating the effect of the stellar mass, with masses as labelled.
	These models assume that the proton superconductivity critical temperature is density independent and very high. 
	See text for details and \protect\fig{fig:Cool_Mass2} for the effect of density dependent $T_{c, p}$.
		}
	\label{fig:Cool_Mass}
\end{figure}

Regarding the neutron star model mass, its effect on the cooling is threefold.
Most dense matter EOS, including the one from \cite{Akmal:1998aa} that we employ, have the effective masses of neutrons and protons 
decreasing with increasing density.
So, the higher the mass of the star, the lower the nucleon effective masses in its inner core. 
From \eq{Eq:eps_nu} one can see that $\epsilon^{nn}_\nu \propto (m^*)^4$, while the specific heat $c_V$ is simply proportional to $m^*$.
Thus, the neutrino cooling function $\ell = \epsilon_\nu / c_V$, which determines the cooling of a neutron star during neutrino cooling
phase (see \cite{Ofengeim:2015daa} and references therein for more details), is proportional to $(m^*)^3$. 
Consequently, $\ell$ decreases when the mass of the star increases, which means that more massive stars cool more slowly.
(This conclusion only holds as long as the more massive star does not have a new neutrino process allowed by its higher central density, in which
case it will cool faster, but we excluded this possibility in our study.
The density dependence of the pairing gaps can also contradict this simple conclusion
as will be illustrated below in \fig {fig:Cool_Mass2}.)
The second effect of increasing the star's mass is the increase in its surface gravity $g_s$ which results in thinner, and hence less insulating, envelope:
for a given internal temperature a more massive star will have a slightly larger surface temperature $T_\mathrm{eff}$.
Finally, the increase in gravity with mass increases the red-shift and results in a decrease in $T_\mathrm{eff}^\infty$ as seen by an observer at infinity.
The total effect of mass dependence is illustrated in \fig{fig:Cool_Mass} showing the decrease of the effective mass with density is the dominant one:
more massive stars have slightly higher $T_\mathrm{eff}^\infty$.
The mass effect is, however, quite small and in our subsequent study we will only consider models of mass 1.4~$\mathrm{M}_\odot$.
Moreover, the small change in $\ell$ with mass is very similar to a change in $R_\sigma$ (see \eq{Eq:eps_nu} and (\ref{Eq:eps_ax}))
which we do fully take into account in our MCMC runs:
our conclusions about $R_\sigma$ hence encompass uncertainties in the star's mass.

\begin{figure}
	\includegraphics[width=1.0\columnwidth]{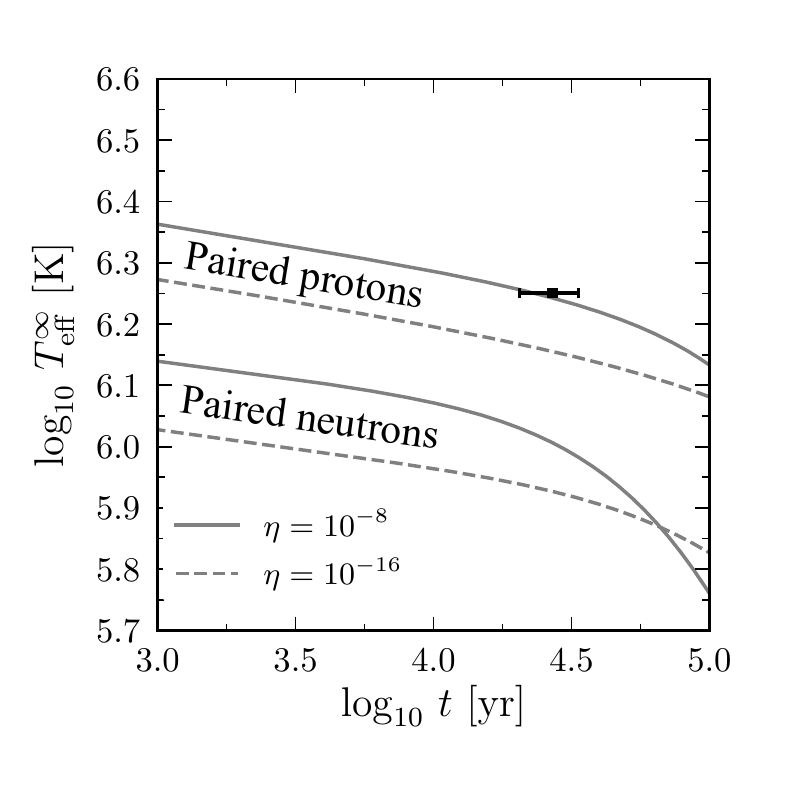}
	\caption{Same as \fig{fig:Cool_crustSF} with four models illustrating the basic effects of proton versus neutron pairing and envelope chemical composition.
	The two upper curves have protons \Singlet-paired in the whole core with a density independent $T_c$ of $7.3\times 10^9$ K and normal neutrons
	while the two lower curves have normal protons and \Triplet-paired neutrons in the whole core with a density independent $T_c$ of  $10^9$~K.
	For each pairing scenario the two curves correspond to two different thickness of carbon layer in the envelope, $\eta$ in \eq{eq:eta}, as labelled.}
	\label{fig:Cool_Tcn_Tcp0}
\end{figure}

As a last step in our characterization of the basic scenario we want to quantitatively evaluate the strength of the argument that core neutron \Triplet superfluidity
is incompatible with the observed high \Te of J1732 \cite{Klochkov:2015rr,Ofengeim:2015daa} and only strong proton superconductivity is acceptable.
We illustrate the basic problem in \fig{fig:Cool_Tcn_Tcp0} by comparing two models with either strong proton \Singlet superconductivity and normal neutrons or
neutron \Triplet superfluidity and no proton superconductivity in the core (but both with neutron \Singlet superfluidity in the inner crust).
In both cases neutrino processes in which the paired component participates are strongly suppressed when $T \ll T_c$, which is the reason for the high \Te 
in the strong proton superconductivity case, but in the neutron superfluidity case there is also a strong flash of neutrinos emitted by the PBF process
when $T \leq T_c$ and result in lower predicted \Te.
The assumed value of \Tcn, $10^9$ K, is higher than the present core temperature of J1732 but not much higher and lead to a very strong cooling
effect from the PBF process; other values will be considered in the MCMC run described below. 
The same \fig{fig:Cool_Tcn_Tcp0} also illustrates the effect of the envelope chemical composition by presenting two models with either $\eta = 10^{-12}$
or $10^{-8}$ for each scenarios: 
a successful model appears to require both strong proton superconductivity (with no neutron superfluidity) and a thick layer of C at the surface.

\begin{figure}
	\includegraphics[width=1.0\columnwidth]{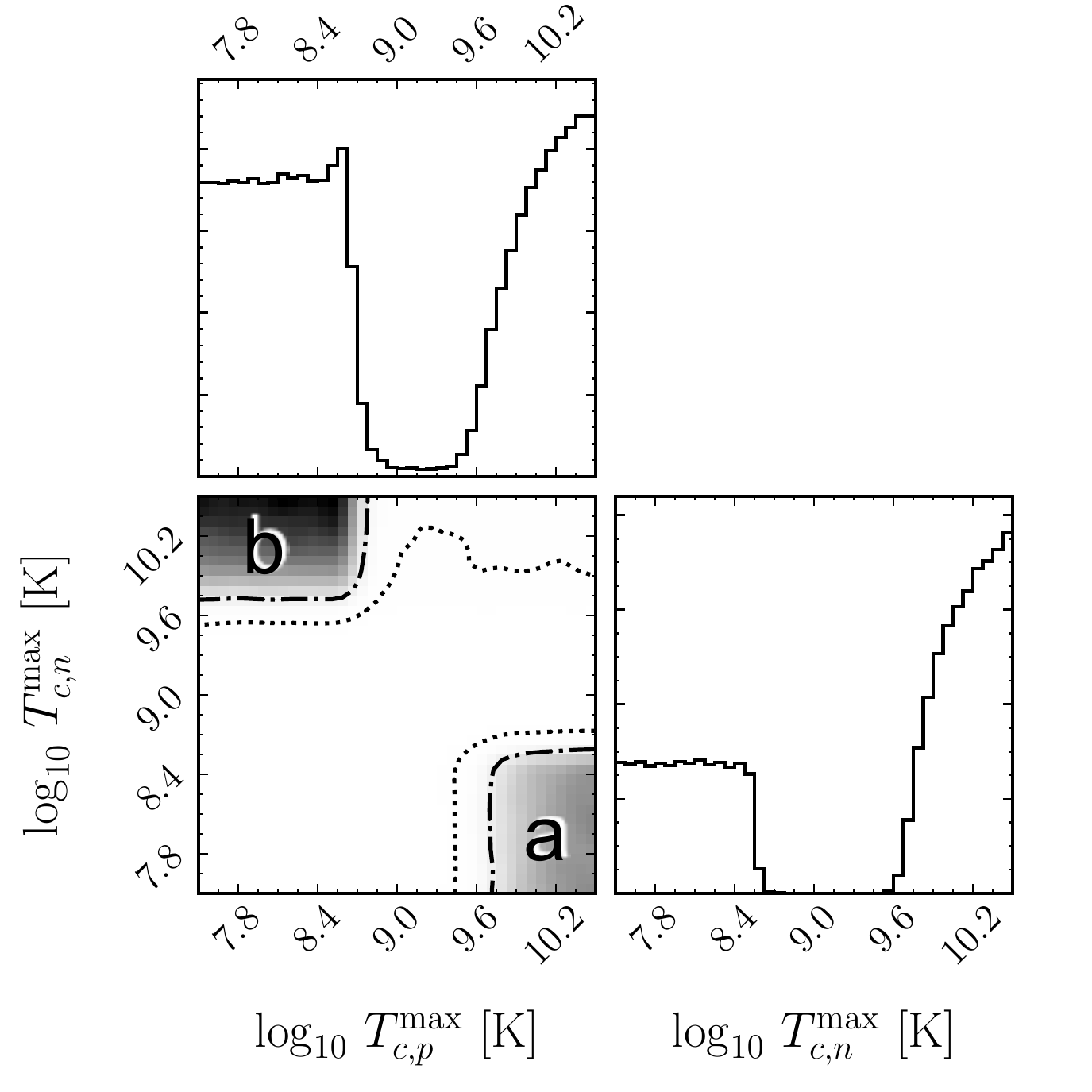}
	\caption{Corner plot of the $\log_{10} T_{c,p}^\mathrm{max}$ and $\log_{10} T_{c,n}^\mathrm{max}$ posterior 
		probability distributions, in the diagonal panels,
         	     and their correlation, in the bottom left panel, from our MCMC run A.
		    Dash-dotted and dotted lines show 90\% and 99\% quantiles respectively, and the grayscale density is proportional to the probability density. 
		    The high probability region marked as ``a'' corresponds to strong proton superconductivity, i.e., with high $T_c$, and
		    weak neutron superfluidity, i.e., with small $T_c$, while the peak marked as ``b'' has the opposite, weak proton superconductivity
		    and strong neutron superfluidity.
		   Given the present core temperature of J1732, peak ``a'' implies normal neutrons and peak ``b'' normal protons in its core.
		See also \protect\fig{fig:Cool_Tcn_Tcp} and text for details.
	            }
	\label{fig:Corner_Tcn_Tcp}
\end{figure}

\begin{figure}
	\includegraphics[width=1.0\columnwidth]{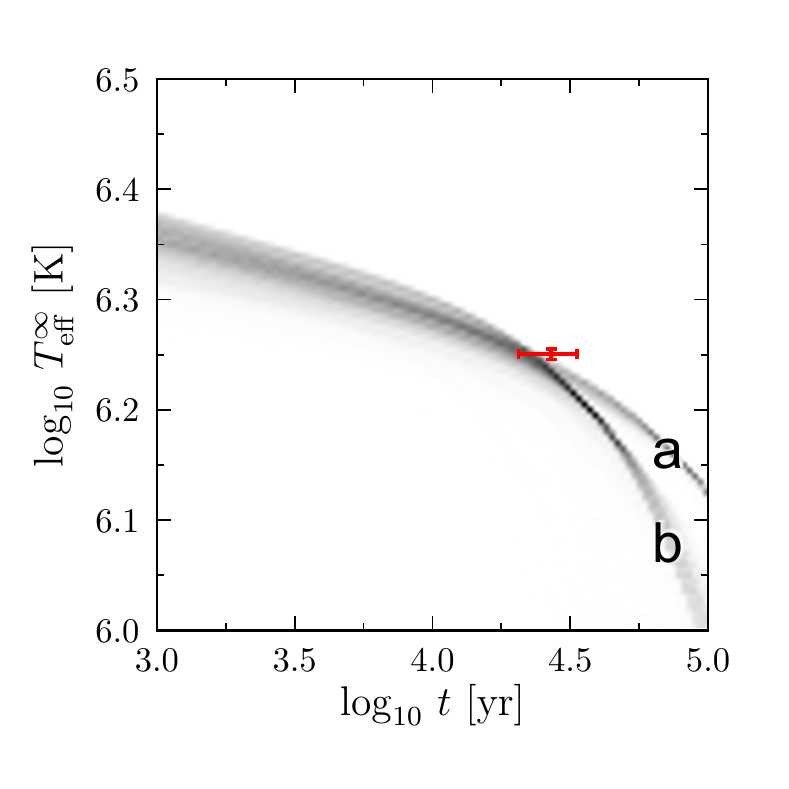}
	\caption{{\em Density of cooling curves} corresponding to \fig{fig:Corner_Tcn_Tcp}.
		10,000 curves were used to generate this density plot, a representative sample of the MCMC run. 
		The two different sub-populations, corresponding to the two separated probability peaks ``a'' and ``b'' seen in \fig{fig:Corner_Tcn_Tcp},
		clearly appear by their cooling behavior at late times.
		The data point corresponds to J1732 with $1\sigma$ uncertainties. 
		See details in the text.
	}
	\label{fig:Cool_Tcn_Tcp}
\end{figure}

To quantify this conclusion we ran an MCMC simulation varying the strengths of both types of pairing and also the amount of C in the envelope $\eta$ (\eq{eq:eta}).
The parameters of our MCMC are four parameters to control the neutron and proton pairing
in the core, their maximum $T_c$ as $T_{c, N}^\mathrm{max}$, and width of the distribution $\Delta k_{F, N}$, (\eq{eq:Tc}), and $\eta$.
Our prior probabilities for these parameters are:
\\
- uniform in $\log_{10} T_{c, N}^\mathrm{max}$ between 7.5 to 10.5 (in K),
\\
- uniform in $\Delta k_{F, N}$ between 0.1 and 1 (in fm$^{-1}$).
\\
- uniform in $\log_{10} \eta$ from $-16$ to $-7$,
\\
the first two applying for $N$ being both $p$ and $n$
(the centers of the two $T_c$ Gaussians are at fixed positions of 2.0 and 0.5 fm$^{-1}$ for $n$ and $p$, respectively, as stipulated previously).
Since the star's mass has little effect we only consider models with a mass of 1.4 \Msun built with the EOS of \cite{Akmal:1998aa} and fix the
inner crust neutron \Singlet superfluidity using the model of \cite{Schwenk:2004aa}.
We also, for now, do not consider the uncertainty on $R_\sigma$ and fix it to its fiducial value of 0.25.
With the possibility of a bimodal distribution between neutron superfluidity versus proton superconductivity we ran 
several parallel tempered chains to properly map the expected separate probability peaks.
We call this our ``MCMC run A''.
Two extracts of the results of this MCMC run are presented on \fig{fig:Corner_Tcn_Tcp} and \ref{fig:Cool_Tcn_Tcp}. 
On \fig{fig:Corner_Tcn_Tcp} we show a corner plot of the $\log_{10} T_{c,p}^\mathrm{max}$ and $\log_{10} T_{c,n}^\mathrm{max}$ posterior probability distributions
and their correlation. 
One can clearly see the bi-modality of the posterior parameter distributions. 
J1732 can be fitted assuming either strong proton superconductivity and normal neutrons (marked as ``a'' in the figure) or strong neutron superfluidity and normal protons (marked as ``b'' in the figure). 
In less than 1\% of cases (see the 99\% quantile) we can have both, but this is actually realized when $\Delta k_{F, p}$ is small and protons are
superconducting only in a very small fraction of the core.
We do not show the distribution of widths, $\Delta k_{F, N}$, because they show nothing interesting except that models require large widths,
i.e., extensive pairing with the sole exception of these less than 1\% cases of simultaneous $n$ and $p$ strong pairing
in which a very small proton pairing width is required.

For both $\log_{10} T_{c,p}^\mathrm{max}$ and $\log_{10} T_{c,n}^\mathrm{max}$ their posterior probability distribution is flat below 8.4:
this simply reflects the fact that, at such low $T_c$, pairing would occur when $T$ is smaller than the present $T$ in the core of J1732 and 
has hence no effect in fitting the data point.
Notice that in the case of \Tcn the value of $10^9$ K as used in \fig{fig:Cool_Tcn_Tcp} is totally excluded along with a wide range of values above and below it.
In the case of \Tcp almost the same range of values is also strongly, but not completely, excluded.
This range of excluded values corresponds to models in which the pairing phase transition would have occurred not too early in the life of the star
but before the age of J1732 and would have resulted in too much neutrino cooling from the PBF process.
The total versus partial exclusion of this range for \Tcn versus \Tcp is due to the fact that the PBF process for the \Triplet neutron gap is much
stronger that for the \Singlet proton gap and thus the former has to be totally excluded.
In contradistinction to this, the upper range of allowed values, for both \Tcp and \Tcn, corresponds to models in which the phase transition occurred 
when the star was very young 
and very hot so that at a still early age $T$ dropped well below $T_c$ and neutrino emission, included the PBF process either from $n$ of $p$ pairing,
became strongly suppressed resulting in a hot star at the present estimated age of J1732.

In \fig{fig:Cool_Tcn_Tcp} we show a plot of  the {\em density of cooling curves} sampled from the MCMC run. The bi-modality is also clearly visible. 
Curves cooling faster after $10^{4.5}$ years have strong neutron superfluidity while the other ones have strong proton superconductivity. 
The difference is due to the specific heat since, the neutron-proton fractions in the core being of the order of 90\%-10\%, suppressing the neutron 
contribution to the specific heat by pairing has a much stronger effect than suppressing the proton contribution.
At earlier ages, before 10$^{4.5}$ years, the two families have very similar, but still slightly distinct, cooling behavior since they have been 
selected by the MCMC to fit the data point.
It is also clear that most curves go below the central value of the data but are passing within less than one $\sigma$ away from it.
The distribution of $\eta$ will be discussed in the next section but large values of $\log_{10} \eta$ are always  strongly preferred.

The dichotomy exhibited in \fig{fig:Corner_Tcn_Tcp} can be resolved comparing the results with theoretical predictions of the size of the
respective proton \Singlet and neutron \Triplet gaps \cite{Sedrakian:2018aa,Page2014}
that predict small neutron \Triplet pairing $T_c$, smaller than $3\times 10^9$ K, and large proton \Singlet $T_c$, much higher than $10^9$ K.
Thus, the peak with $\log_{10} T_{c,p}^\mathrm{max} > 9.5$ and $\log_{10} T_{c,n}^\mathrm{max}<9$ fits well within theoretical expectations while 
the other one with $\log_{10} T_{c,p}^\mathrm{max} < 9$ and $\log_{10} T_{c,n}^\mathrm{max}>9.5$ is supported by hardly any theoretical model.
We will, hence, work within the same framework as \cite{Klochkov:2015rr} and \cite{Ofengeim:2015daa} and conclude that the high observed
\Te of J1732 requires strong proton \Singlet superconductivity and excludes neutron \Triplet superfluidity with high or moderate ($\gtrsim 3 \times 10^8$ K) $T_c$.
Strong proton superconductivity mean that the corresponding critical temperature $T_c$ is much larger than $10^9$ 
in most of the star's core.
When comparing with theoretical models of proton pairing (see, e.g., Figure 21.11 in \cite{Page2014})
this requirement implies that the mass of the neutron star J1732 is low;
how low is model dependent but very unlikely higher than, and more likely well below, 1.4~$\mathrm{M}_\odot$.
We illustrate this in \fig{fig:Cool_Mass2} using the CDDK model \cite{Chen:1993fk} for the proton \Singlet gap.
For this model, there is a strong increase in neutrino emissivity at high densities (see \fig {fig:Gap_nu_axion})
and a neutron star mass well below 1.4~\Msun\,is strongly favored.
Actual mass values deduced from such a study are extremely dependent on the assumed density profile of the proton gap.
Among the many published theoretical prediction of the proton \Singlet gap this CCDK model is one of the few that reach high densities:
most other published gap would imply much lower masses.

\begin{figure}
	\includegraphics[width=1.0\columnwidth]{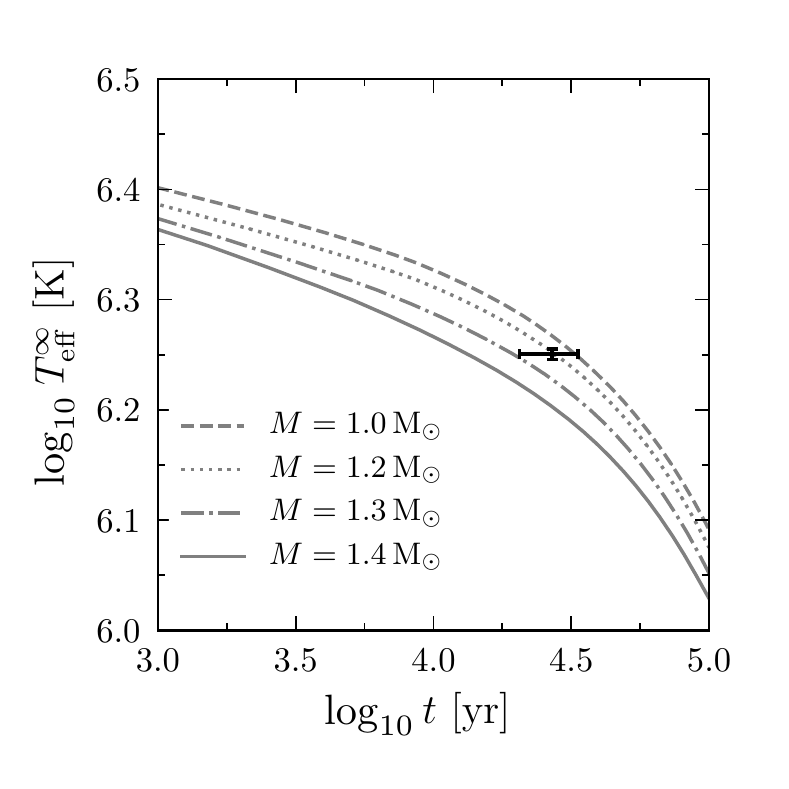}
	\caption{Same as \fig{fig:Cool_Mass} with four models illustrating the effect of the stellar mass, with masses as labelled, when a density dependent proton \Singlet gap, model CCDK from \protect\cite{Chen:1993fk}, is used.
		See text for details.}
	\label{fig:Cool_Mass2}
\end{figure}

\section{Improved constraints on Axion-like particles}
\label{sec:axion_constraint}

\begin{figure*}

  \includegraphics[width=1.0\columnwidth]{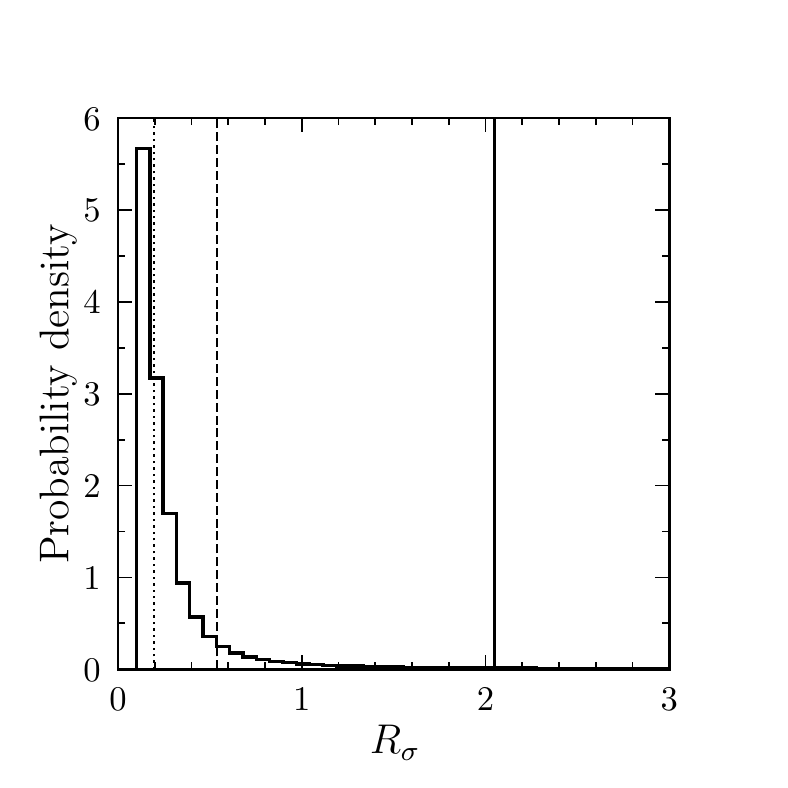}
  \includegraphics[width=1.0\columnwidth]{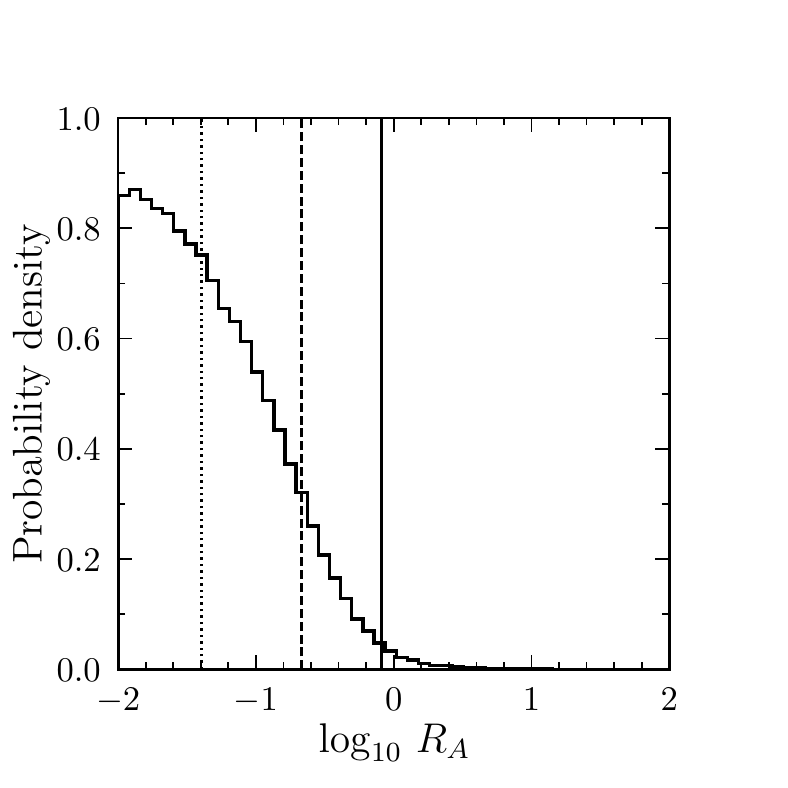}
  \\
  \includegraphics[width=1.0\columnwidth]{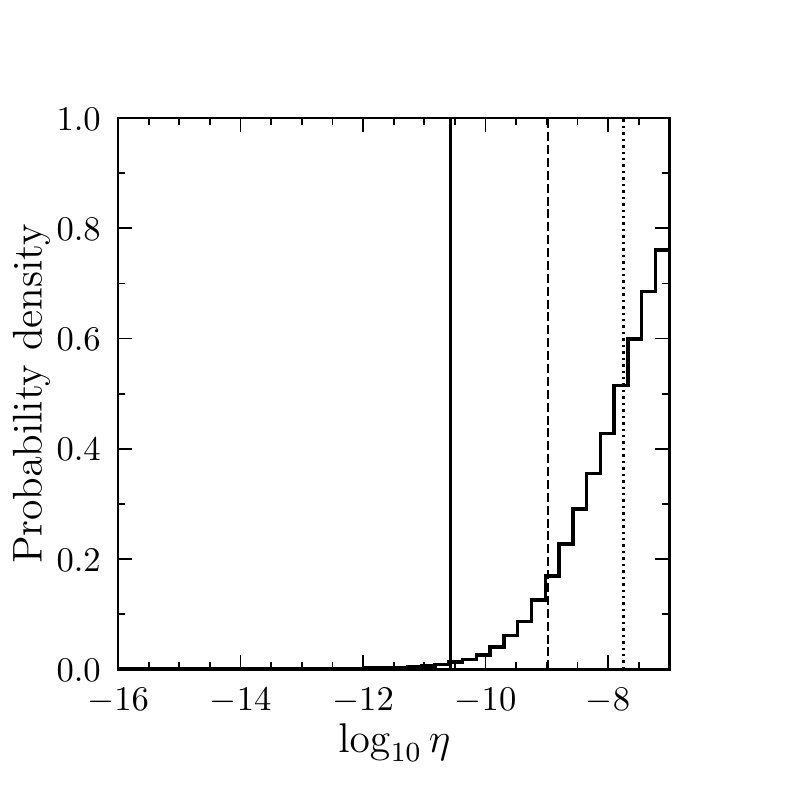}
  \includegraphics[width=1.0\columnwidth]{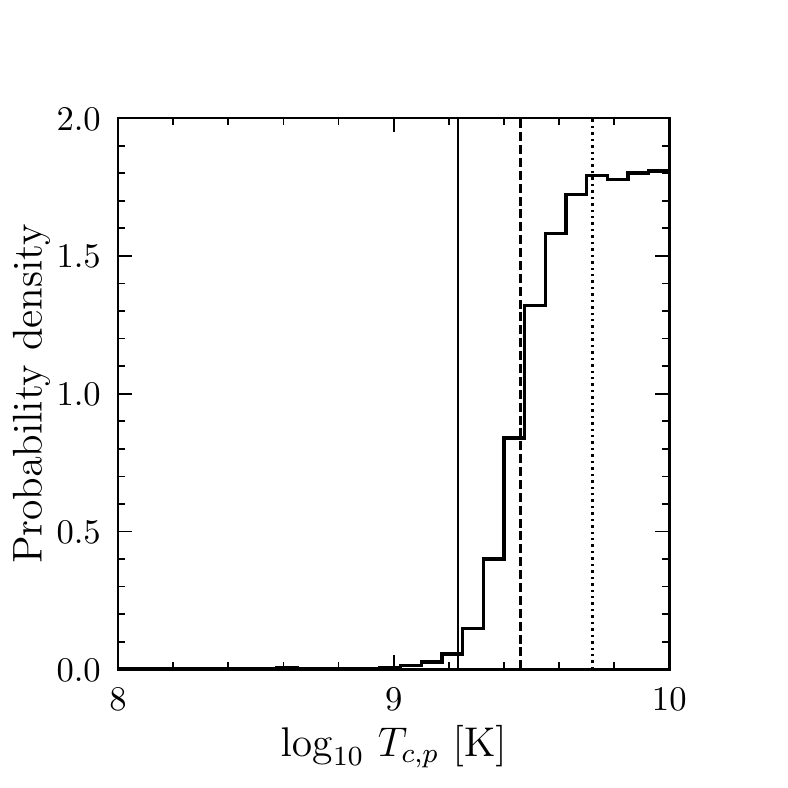}
  \caption{Histograms of posterior probability distribution of $R_\sigma$, $\log_{10} R_A$, $\log_{10} \eta$, and $\log_{10} T_{c,p}$ 
  from our MCMC run B.
  Vertical lines are: dotted = 50\%, dashed = 90\% and solid = 99\% quantiles.
  See text for details.}
  \label{fig:Histograms}
\end{figure*}

Having characterized the basic scenario and the requirements imposed by the observed high \Te of J1732 we are now able to carefully assess
the efficiency of neutrino cooling and constrain axion emission. For this purpose we extend our previous MCMC run by adding both $R_\sigma$ and $R_A$, see \eq{Eq:Rs_RA}, as parameters.
We now consider 5 parameters and their prior distributions are uniform in natural or logarithmic scale:
\\
- $\Rs$: $0.1\leq \Rs \leq 3$
\\
- $R_A$: $-2 \leq \log_{10} R_A \leq 2$
\\
- $\eta$: $-16 \leq \log_{10} \eta \leq -7$
\\
- $T_{c,p}$: $8 \leq \log_{10} T_{c,p} [\mathrm{K}] \leq 10$~(density independent)
\\	
- $T_{c,n}$: $8 \leq \log_{10} T_{c,n} [\mathrm{K}] \leq 9$~~(density independent).
\\
As it was explained in the previous section, we have chosen the high $T_{c,p}$ -- low $T_{c,n}$ mode from the bi-modal distribution 
(see \fig{fig:Corner_Tcn_Tcp}). 
In this case $T_{c,n}$ was lower than the core temperature of J1732 at present age so that neutrons were normal and this gap had no effect
on fitting the J1732 data point, but we still include it in the present calculation due to its large potential effect on the neutrino emission.
Neutron \Singlet pairing from \cite{Schwenk:2003vq} is present in the inner crust and the star's mass is fixed at $M = 1.4$ M$_\odot$.
This is our ``MCMC run B''.

The posterior distributions of 4 relevant parameters are presented in \fig{fig:Histograms}.
As in the cases studied above we obtain a strong lower limit on $T_{c,p}$ that is needed to suppress neutrino emission by the modified Urca process
and a highly conductive envelope, i.e., an as thick as possible carbon layer, to raise as much as possible the observable \Te.
In line with the difficulty to keep a neutron star so hot for such a long time, see, e.g., \fig{fig:Cool_Tcn_Tcp}, allowing for a reduction of $\epsilon_\nu^{nn}$
through $R_\sigma$ favor small values of $R_\sigma$ (the median of \Rs distribition is $\approx 0.2$).
Similarly, small values of $R_A$ are strongly preferred but with a rather broad distribution.
Notice that the constraint we obtain on $R_\sigma$ is more drastic than the one on $R_A$ because $R_\sigma$ controls both $\epsilon_\nu^{nn}$
and $\epsilon_a^{nn}$.
The resulting constraint on $T_{c,n}$ is not shown since it simply consists in keeping neutron normal, i.e., $T_{c,n}$ must be lower than the present
internal temperature of J1732.

\begin{figure}
  \includegraphics[width=1.0\columnwidth]{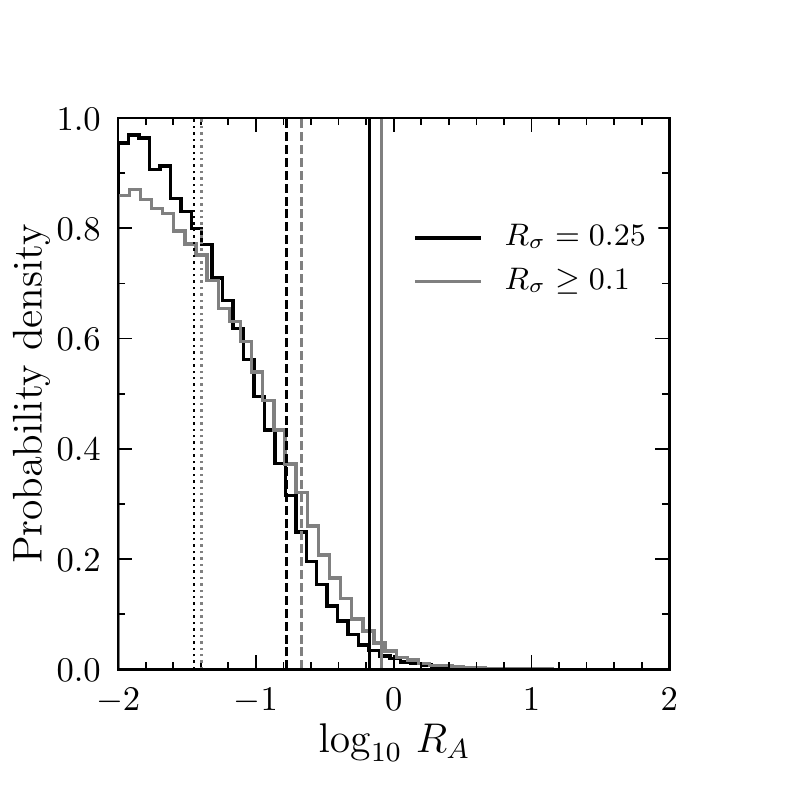}
  \caption{Comparison of posterior $R_A$ probability distributions with variable, MCMC run B, and fixed, MCMC run C, $\Rs$. 
  	      Vertical lines are the same quantiles, 50\%, 90\% and 99\%, as in \fig{fig:Histograms}.
              Distribution with variable $\Rs$ (gray line) is identical to one in \fig{fig:Histograms}. 
              Black lines corresponds to distribution of $R_A$ obtained with fixed $\Rs = 0.25$. 
              See details in the text.}
  \label{fig:Histograms_RA}
\end{figure}

\begin{figure*}
  \includegraphics[width=1.0\columnwidth]{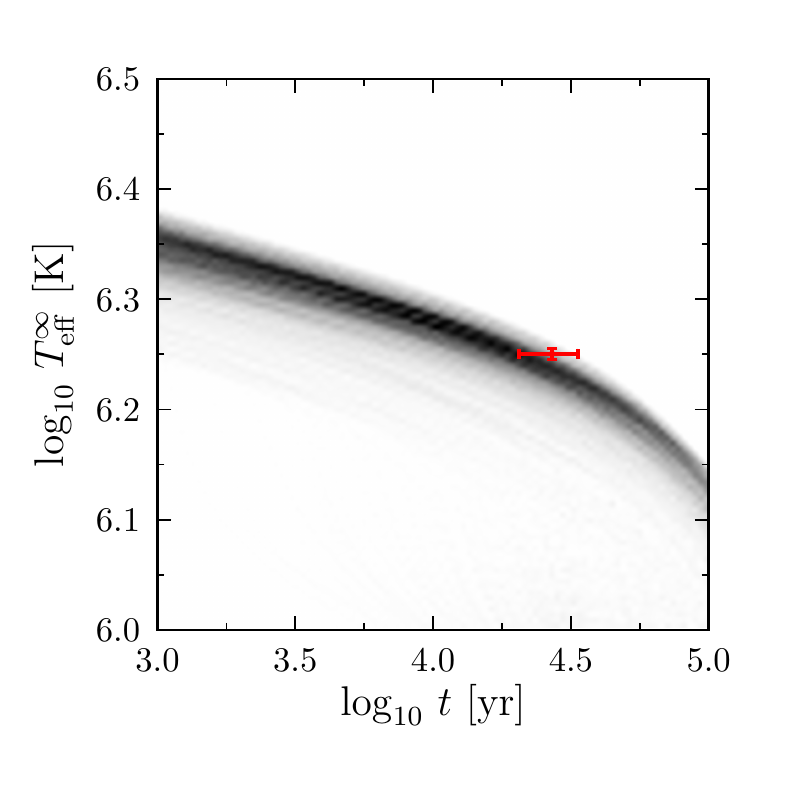}
  \includegraphics[width=1.0\columnwidth]{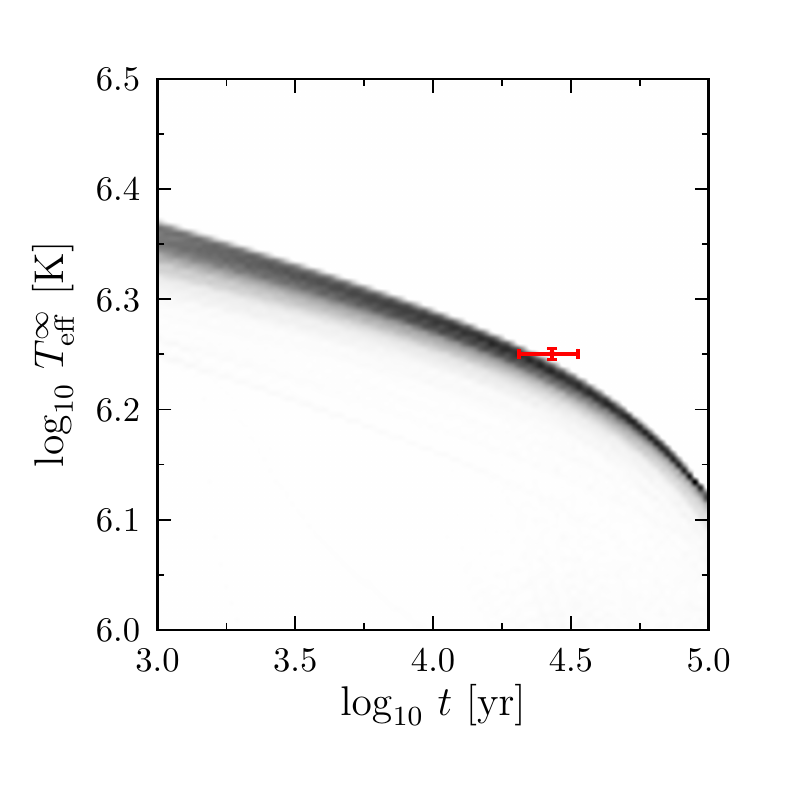}
  \caption{Density of cooling curve to illustrate of the role of \Rs: left panel has $0.1 \leq \Rs \leq 3$ and right panel has a fixed \Rs = 0.25.
  		10,000 curves were used to generate each density plot, a representative sample of the MCMC runs B and C.
		The data point corresponds to J1732 with $1\sigma$ uncertainties. 
  }
  \label{fig:Cool_Rs}
\end{figure*}

We also made a comparative MCMC run with a fixed value $\Rs = 0.25$,
our fiducial value for \Rs.
Other parameters of the run are identical and this is our ``MCMC run C".
The resulting comparison of $R_A$ distributions is presented in \fig{fig:Histograms_RA} which shows that the posterior distribution of $R_A$
changes only slightly whether $\Rs$  is variable or fixed.
It is also rather insensitive to changes in neutron star mass and to the presence or absence of neutron superfluidity in the crust (as shown in the previous subsection). 
Thus, the $R_A$ distribution results to be remarkably robust.

To  illustrate further the impact of $\Rs$, let us look at the cooling curves density plot, similar to \fig{fig:Cool_Tcn_Tcp},
in \fig{fig:Cool_Rs} which shows the comparison of variable $\Rs$ case with fixed  $\Rs$ case. 
Several interesting and illustrative points can be made from this plot. 
First, the variable $\Rs$ run results in a wider distribution of cooling curves. 
This is to be expected: the more degrees of freedom we have, the wider the resulting distributions will be
and is a natural payoff from increasing the number of parameters. 
On the other hand, we can fit the data point  better with more parameters, which can be seen by comparing left and right panel of the figure. 
Second conclusion is that it is remarkably difficult to fit J1732 with a cooling curve. 
Even with variable $\Rs$ most of the curves go below the data point (inside $1\sigma$ uncertainty region, but still below the point itself). 
It is important to note that there are curves that go exactly through the data point 
but the number of such models is very small and they are barely visible on a density plot, especially on the right panel. 
One has to keep in mind that MCMC not only finds the best fit, but also ``walks" around it;
here we can see that  it can easily walk below the data point, but it is very difficult to go above it. 

From \fig{fig:Histograms_RA} we can infer that the 90\% upper limit on $\log_{10} R_A$ is $-0.67$, which means that  90\% confidence 
$g^2_{ann} < 7.7 \times 10^{-20}$ -- a factor of 5 improvement over the SN 1987A limit.  
Additionally, we claim that our estimate is more robust to changes
in the physical model as demonstrated in the previous sections. 
We made a bootstrap estimation of the standard deviation of the 90\% quantile which turned out to be $\sim 4\%$.
Moreover, taking into account variations caused by changes in the stellar model mass and $\Rs$ we estimate at most a 10\%
uncertainty and can state that the 90\% upper limit  on $\log_{10} R_A$ is $-0.67 \pm 0.06$.

\begin{figure*}
  \includegraphics[width=2.0\columnwidth]{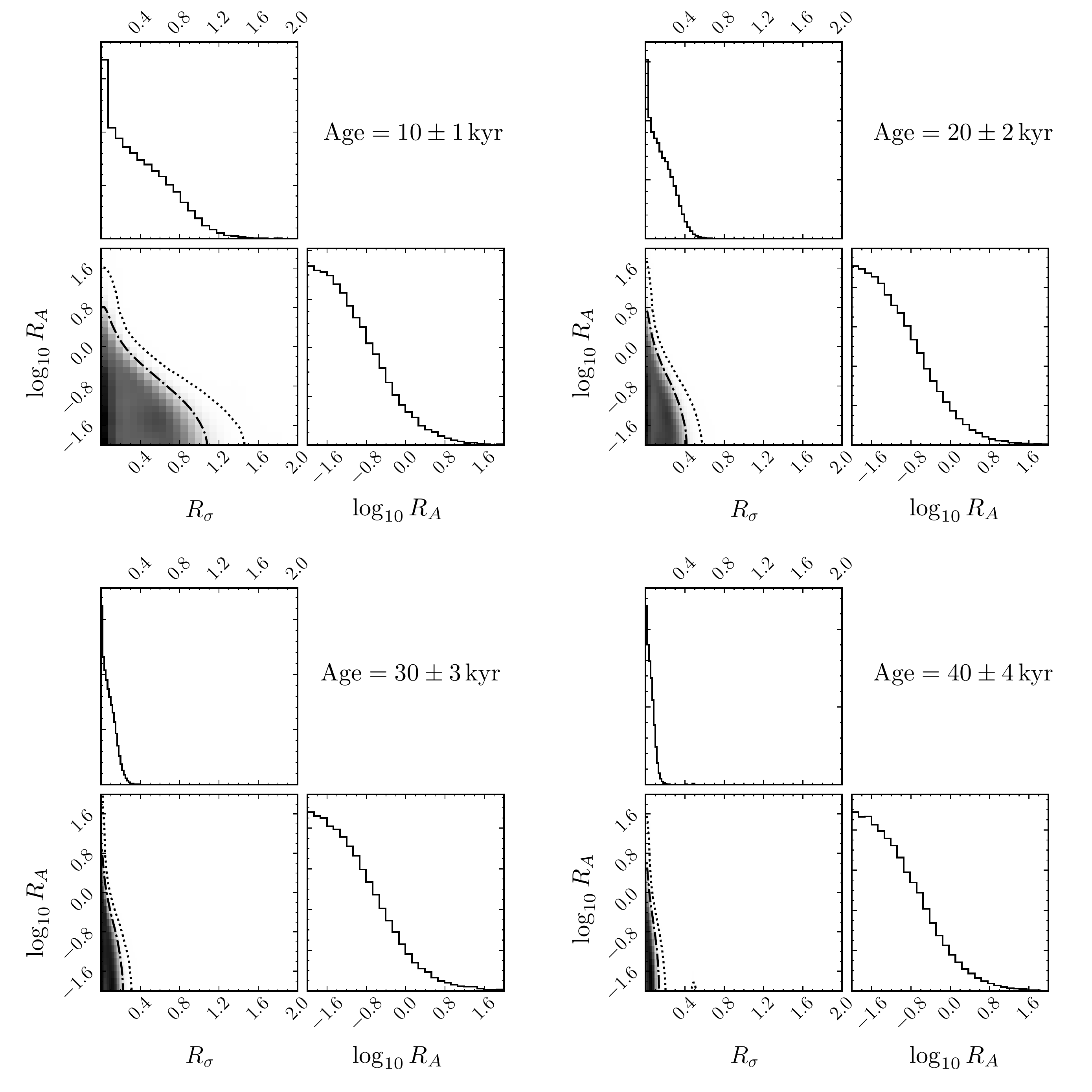}
  \caption{Corner plots of posterior probability distributions of $R_\sigma$ and $R_A$ for our four MCMC runs D1, D2, D3, and D4,
  		assuming different ages for the remnant J1732, as indicated in the panels corners.
		See text for details
  }
  \label{fig:Corner_Age}
\end{figure*}

Thus far our analysis was based on the age of J1732 given by  $27 \pm 6.5$ kyrs.
We now enquire how our constraints would change if a better age estimate with much smaller uncertainty were to become available. 
We considered four cases of age 10, 20, 30, and 40 kyrs with, in each case a small 10\% uncertainty and ran four MCMC runs D1, D2, D3, and D4, respectively.
The results are presented in \fig{fig:Corner_Age} four corner plots for the resulting distribution of both $R_\sigma$ and $R_A$.
These runs used the same parameters and condition as run B with the one important change beside the star's age --  
we relaxed the condition on $R_\sigma$ extending its range to:
\\
- $\Rs$: $0.0< \Rs \leq 3$.
\\
Allowing very small values of $R_\sigma$ may be unrealistic but allows us to obtain reasonable fits for large ages. The resulting distribution of $\eta$ and $\log_{10} T_{c,N}$ are practically the same as in previous runs and not shown. As can be expected from the results of previous runs we obtain very strong constraints on $R_\sigma$ since it controls 
both $\epsilon_\nu^{nn}$ and $\epsilon_a^{nn}$, and the older the remnant the smaller $R_\sigma$ has to be. The distributions of $R_A$ are almost identical in all four cases, again a consequence of the dominant effect of $R_\sigma$. They are, however, extending to larger values of $R_A$ simply because we arbitrarily allowed $R_\sigma$ to be vanishingly small.
Nevertheless, in the case that an old age of J1732 could be inferred in the future and a strong theoretical lower limit on the value of $R_\sigma$
could be obtained then we would be able to constraint the axion-neutron coupling constant much more significantly than presently possible.

Finally, it is important to emphasize here that our ability to obtain meaningful constraints on axion coupling by means of studying J1732 relies on its high surface temperature. 
We employed the value $T_\mathrm{eff} = 2.24$~MK (corresponding to $T_\mathrm{eff}^\infty =1.78$~MK) from Ref. \cite{Klochkov:2015rr}. 
However, in Ref. \cite{SKPW17} the authors investigate the possibility that J1732 has a non-uniformly emitting surface (i.e., hot spots). 
In this case the cold  component (corresponding to most of the stellar surface) has an effective local temperature of $T_\mathrm{eff} = 2.04$~MK which is lower than the value we used. 
A lower surface temperature means a less restricting constraint on axion coupling. 
However, the authors of Ref. \cite{SKPW17} have found that this scenario is unlikely for several reasons (yet cannot be ruled out as the spectral fits are ``formally acceptable'').

\section{Discussion of Related Works}
\label{sec:discussion}

Beside the previously mentioned constraints on $g^2_{ann} + g^2_{app}$ \cite{Koshiba1992,Raffelt:2006cw} from SN 1987A,
the cooling of three middle-aged ($10^5 - 5 \times 10^5$ yrs) pulsars, PSR 0656+14, PSR 1055-52, and PSR 0633+1748 
(``Geminga''), which all are in the photon cooling era, was studied in \cite{Umeda:1998fk} and upper limits on the axion mass and $f_a$ obtained:
as in our study, a too strong axion-nucleon coupling would have lead, during the previous neutrino cooling phase of these three stars,
to a more rapid cooling resulting in temperatures during the photon-cooling era that are lower than observed.
Interesting constraints were obtained, in the best case $g^2_{ann}  < 1.6 \times 10^{-19}$ within the KSVZ model when neutrons are not paired.
However, the robustness of this result is difficult, if not impossible, to assess for several reasons.
Firstly, there are, unfortunately, no realistic atmosphere models, which require strong magnetic fields for such pulsars, 
that can fit the observed soft X-ray spectra of these pulsars and, hence, their effective
temperatures are only estimated using a blackbody spectrum and an error of at least a factor of two is very possible.
Not being associated with a supernova remnant implies that their age can only be estimated from a simple pulsar braking law:
the error could be small or very large and there is no known way to reliably assess it.
Finally, at such large ages, heating processes  (see, e.g., \cite{Umeda:1993aa}) coupled with uncertainty about
the envelope chemical composition (see, e.g., \cite{Page:1997aa}) may seriously alter their thermal evolution
again in a way that is practically impossible to assess.

The possible observation of rapid cooling in the young neutron star ``Cas A'' in the supernova remnant Cassiopea A \cite{Heinke:2010aa}
has been interpreted as due to the emission of 
a strong burst of neutrinos from the PBF reactions involving Cooper pairs of neutrons in the spin triplet \Triplet channel 
during the onset of the phase transition \cite{Shternin:2010qi,Page:2011aa}. 
This process would also produce axions and an earlier study found that the resulting cooling would be incompatible with observations for $g^2_{ann} > 2 \times 10^{-19}$ \cite{Sedrakian:2015krq}. However, and perhaps more interestingly for $g^2_{ann} \simeq 1.4 \times 10^{-19}$ axion cooling provided a good fit to the data \cite{Leinson:2014ioa}. Note that this coupling strength is a bit smaller than and compatible with the SN1987A constraint in Eq.~\ref{eq:sn1987a}.  Does this hint at the possible existence of axions? Lacking a thorough analysis of systematic errors that can arise from a myriad of poorly 
understood microphysical input physics that play a role in neutron star cooling models it would be prudent to be skeptical,
but one cannot simply dismiss the possibility either. However, since this value of $g^2_{ann}$ is about a factor of 2 larger than the limit we have derived it would be incompatible with our interpretation of J1732. 

The tension between our results, particularly about the size of the neutron \Triplet gap, and the above mentioned interpretation of the
rapid cooling of Cas A is possibly even stronger.
We inferred that $T_{c, n}^\mathrm{max}$ must be {\em smaller} than $\sim 2.5 \times 10^8$ K, given the present core temperature of J1732,
while all three works \cite{Shternin:2010qi,Page:2011aa,Leinson:2014ioa} on the rapid cooling of Cas A  imply that $T_{c, n}^\mathrm{max}$ 
must be {\em larger} than $\sim 5 \times 10^8$ K, limited by the core temperature of Cas A.
It is important to consider that these deduced values of $T_{c, n}^\mathrm{max}$ refer to the maximum value of $T_{c, n}$
in the neutron star J1732 and the neutron star Cas A, respectively.
In our simple Gaussian model of \eq{eq:Tc} and our fixed value of $k_{F, n}^\mathrm{max} = 2.0 $ fm$^{-1}$ in all neutron stars
the maximum value of $T_{c, n}$ is the same and is $T_{c, n}^\mathrm{max}$ since they all have a $k_{F, n}$ in their center that
exceed 2.0 fm$^{-1}$ and thus reach beyond the peak of the Gaussian\footnote{With APR EOS, $k_{F, n}$ already reaches 2.3 fm$^{-1}$
at the center of a 1.0 \Msun star.}.
If, instead, we would have a much larger value of $k_{F, n}^\mathrm{max}$, the peak of the Gaussian could be beyond the central value
of low mass stars: neutron stars of not too high mass would have a maximum $T_{c, n}$ reached at their center and
stars of increasing mass would have a increasing maximum $T_{c, n}$ (until the star is massive enough that its central density reaches the
peak of the $T_{c, n}$ curve).
Such a density dependence of the neutron \Triplet gap has been proposed in \cite{Gusakov:2004fk} in their version of the ``minimal'' cooling paradigm.
If this were the case then the above models for the rapid cooling of Cas A and our analysis of J1732 could be compatible and 
simply imply that the Cas A neutron star is heavier than the J1732 neutron star,
an issue which nevertheless deserves more work but is beyond the goal of the present paper.
There are, however, other possible interpretation of the cooling of Cas A, e.g., 
a quark color-superconducting phase transition in a massive star \cite{Sedrakian:2013aa} instead of neutron superfluidity, or 
delayed core thermal relaxation because of a low core conductivity \cite{Blaschke:2012aa}, 
among the many proposed alternatives, which are not at odds with our present results.
Another possibility would simply be that the Cas A neutron star claimed ``rapid cooling'' is an instrumental artifact and thus not real \cite{Posselt:2013aa}.

In a wider setting, novel constraints on the size of the nucleon pairing gaps were recently explored in \cite{BHSP18} which also worked
within the minimal cooling paradigm of \cite{Page:2004fy} making more quantitative several of the conclusion of  \cite{Page:2009aa}.
In the three families of scenarios considered in \cite{BHSP18} the first two explicitly excluded the ``carbon stars'', i.e. both Cas A and
J1732 whose best temperature measurements employed carbon atmosphere models, while the third one included them and the results
about the proton \Singlet and neutron \Triplet $T_c$ curves are illustrated in the figure 4 of that work.
The values of $T_{c, p}^\mathrm{max}$ and $T_{c, n}^\mathrm{max}$ which can be read from that figure are in a very good 
agreement with the high probability peak labelled as ``b'' in our  \fig{fig:Corner_Tcn_Tcp}, even though that work found acceptable
models within a slightly larger range of values because they fitted their cooling models
to a large number of neutrons star while we restricted ourselves to a single object.
However, inspired by the study of \cite{Klochkov:2015rr} and \cite{Ofengeim:2015daa} we searched for the possibility of
a bimodal distribution and used tempered chains in our MCMC runs which effectively found two clear peaks, ``a'' and ``b'' shown in \fig{fig:Corner_Tcn_Tcp}.
We argued in \S~\ref{sec:basic} that the peak ``a'', appears more natural when compared with theoretical predictions and we consider
the $T_c$ values corresponding to the peak ``b'', strong superfluidity with weak superconductivity, as unrealistic.

\section{Conclusion}
\label{sec:conclusion}

Assuming that the thermal evolution of J1732 is dominated by neutrino cooling due to bremsstrahlung reactions involving normal neutrons
in the core we have derived a new constraint on the axion-neutron coupling,  
 $g^2_{ann} < 7.7 \times 10^{-20}$ at 90\% confidence,which is 5 times better than the previously existing constraint from SN 1987A.
For the KSVZ-axion model where $c_n=0.02$, 
at $90$\% confidence we claim that $f_a >6.7 \times 10^7$ GeV, while for the DFSZ-axion model with $c_n=0.5$ we claim that  $f_a >1.7 \times 10^9$ GeV.  
We have discussed how variation of key model parameters and uncertainties associated with the age of the J1732 influence our 
constraint. The high temperature encountered in J1732 not only disfavors additional cooling due to axions but also favors a suppression of the standard neutrino cooling rate. Interestingly, this suppression can be accounted for by an improved treatment of the nucleon-nucleon interaction in the calculation of the bremsstrahlung rate discussed many years ago \cite{Hanhart:2000ae}. 

The widely used constraint from the neutrino signal from SN1987A and the neutron star cooling constraint we discuss here are both subject to uncertainties associated with the phase structure and linear response properties of matter at supra-nuclear density but there are important distinctions that are worth mentioning. In the supernova context, the binding energy emitted 
as neutrinos is well understood but neutrino transport and thermal evolution of the PNS depends on a host of thermodynamic and response properties of hot and dense matter 
that can influence the duration of the neutrino signal \cite{Fischer:2016cyd}. In contrast since J1732 is only compatible with a slow neutrino cooling scenario which implicates 
a specific reaction to play a dominant role, the neutron star cooling model is already constrained rather well. We believe that unless large internal magnetic field decay or some other 
more mysterious source of heating can account for the high temperatures observed in J1732, our constraint for the axion-neutron coupling is robust. 
For models in which the axion coupling to neutrons is not too small, i.e., $c_n\simeq 1$ we conclude that the axion-decay constant  $f_a > 10^9$ GeV.        

As a by product of our study we also found new important constraints on the size of both the proton \Singlet and neutron \Triplet gaps at supranuclear densities
which are exhibited in \fig{fig:Corner_Tcn_Tcp}:
in the J1732 neutron star, $T_c$ for proton \Singlet pairing must be, in most of the core, larger than $4 \times 10^9$ K while
for neutron \Triplet it must be, everywhere in the core, below $3 \times 10^8$ K.
How much these two constraint can be extended to nucleon pairing in the core of other neutron stars depends on density dependence of these nucleon gaps
as well as on the mass of J1732.
Comparison with theoretical prediction of these gaps seem to favor a low mass for the J1732 neutron star.


\begin{acknowledgments}
 MB is fully and DP partially supported by the Consejo Nacional de Ciencia y Tecnolog{\'\i}a with a CB-2014-1 grant $\#$240512. 
 The work of SR was supported by the U.S. DOE under Grants No. DE-FG02- 00ER41132 and in part by the National Science Foundation 
 Grant No. PHY- 1430152 (JINA Center for the Evolution of the Elements). The work of ER was supported in part by the Natural Sciences and Engineering Research Council (NSERC) of Canada, the Canada
Foundation for Innovation (CFI), and the Early Researcher Award (ERA) program of the Ontario Ministry
of Research, Innovation and Science.
MCMC calculations were ran on the {\tt Atocatl} cluster at the Instituto de Astronom\'ia, UNAM.
\end{acknowledgments}

\end{document}